\documentclass[a4paper,11pt]{article}
\usepackage{a4wide,amsmath,amssymb,bbm}
\usepackage[english]{babel}

\usepackage{color}

\usepackage{hyperref,enumerate}
\hypersetup{
    colorlinks,
    linkcolor={black},
    citecolor={black},
    urlcolor={black}
}

\usepackage{mathrsfs}

\newcommand{\tr}{\mathrm{tr}}
\newcommand{\Tr}{\mathrm{Tr}}
\newcommand{\Ad}{\mathrm{Ad}}
\newcommand{\be}{\begin{equation}}
\newcommand{\ee}{\end{equation}}
\newcommand{\alg}{\mathfrak}

\newcommand{\sL}{\mbox{\tiny L}}
\newcommand{\sR}{\mbox{\tiny R}}

\makeatletter

\@addtoreset{equation}{section} \makeatother

\begin{document}

\hfill{NORDITA 2018-122}

\vspace{30pt}

\begin{center}
{\huge{\bf Marginal deformations of WZW models\\  and the classical Yang-Baxter equation}}

\vspace{80pt}

Riccardo Borsato$\, ^{a,b}$ \ \ and \ \ Linus Wulff$\, ^c$

\vspace{15pt}

{
\small {$^a$\it 
Instituto Galego de F\'isica de Altas Enerx\'ias (IGFAE), Universidade de  Santiago de Compostela, Spain}\\
\vspace{5pt}
\small {$^b$\it Nordita, Stockholm University and KTH Royal Institute of Technology, Stockholm, Sweden}\\
\vspace{5pt}
\small {$^c$\it Department of Theoretical Physics and Astrophysics, Masaryk University, 611 37 Brno, Czech Republic}
\\
\vspace{12pt}
\texttt{riccardo.borsato@usc.es, wulff@physics.muni.cz}}\\

\vspace{100pt}

{\bf Abstract}
\end{center}
\noindent
We show how so-called Yang-Baxter (YB) deformations of sigma models, based on an $R$-matrix solving the classical Yang-Baxter equation (CYBE), give rise to marginal current-current deformations when applied to the Wess-Zumino-Witten (WZW) model. For non-compact groups these marginal deformations are more general than the ones usually considered, since they can involve a non-abelian current subalgebra. We classify such deformations of the $AdS_3\times S^3$ string.

\pagebreak 
\tableofcontents

\setcounter{page}{1}


\section{Introduction}
Conformal field theories (CFTs) in two dimensions are of interest for various areas of physics, from condensed matter physics to string theory. In string theory they naturally arise on the worldsheet of the string. In the context of holographic duality, certain two-dimensional CFTs are also known to be \emph{dual to} string theories on three-dimensional anti de Sitter spacetimes~\cite{Maldacena:1997re,Giveon:1998ns,Kutasov:1999xu}. An important instance of the $AdS_3/CFT_2$ duality is obtained by studying string theory on $AdS_3\times S^3\times T^4$. In the case of pure NSNS flux the string is described by a Wess-Zumino-Witten (WZW) model on the group $F=SL(2,\mathbb R)\times SU(2)$, see~\cite{Maldacena:2000hw,Maldacena:2000kv,Maldacena:2001km} and references there. We will be interested mainly in this setup. The CFT description of the worldsheet theory allows to make precise statements about the AdS/CFT duality in this case. A recent example is the duality between the symmetric product orbifold CFT and the string WZW model at level $k=1$~\cite{Eberhardt:2018ouy}.

Generally speaking it is interesting to understand the conformal manifold of a CFT, i.e. the space of marginal deformations generated by adding a local perturbation to the Lagrangian. When applied to the WZW model under study, the marginal deformations correspond to deformations of the supergravity background. They give us (at least in principle) a way to go beyond the usual $AdS_3/CFT_2$ duality and extend it to cases in which e.g. the supersymmetry or the conformal symmetry of the dual $CFT_2$ are broken. Local marginal deformations of WZW models were studied by Chaudhuri and Schwartz (CS)~\cite{Chaudhuri:1988qb}. They found that {a necessary condition for} local operators constructed out of the chiral and antichiral currents as 
\begin{equation}
O(\sigma,\bar \sigma)= c^{a b}J_a(\sigma)\bar J_{ b}(\bar \sigma)\,,
\end{equation}
with $c^{ab}$ some constant coefficients, {to give a marginal deformation is that $c^{ab}$ satisfy}
\be\label{eq:CS-weak-intro}
C\cdot C+ \bar C\cdot \bar C=0\,,
\ee
where we have defined $C^{abc}\equiv c^{da}c^{eb}f_{de}^{\ c}$, and $\bar C^{abc}\equiv c^{ad}c^{be} f_{de}^{\ c}$, and the product is obtained using the Killing metric $K_{ab}$, e.g. $C\cdot C\equiv C^{abc}C^{def}  K_{ad} K_{be}K_{cf}$.
Equation~\eqref{eq:CS-weak-intro} is quartic in $c^{ab}$, and it involves also the structure constants of the algebra of $F$, the Lie group of the WZW model. We will call it the \emph{weak} CS condition. CS were interested in the case of CFTs where the group $F$ is compact. In that case~\eqref{eq:CS-weak-intro} reduces to 
\be\label{eq:CS-strong-intro}
C^{abc}=0\,,\qquad\text{and}\qquad
\bar C^{abc}=0\,,
\ee
which is an equation quadratic in $c^{ab}$ that we will call the \emph{strong} CS condition. It imposes, in fact, a stronger constraint, since it is only in the case of compact groups that it is equivalent to~\eqref{eq:CS-weak-intro}. CS also showed that solutions of the strong condition correspond to abelian subalgebras of $Lie(F)$, since when~\eqref{eq:CS-strong-intro} holds it is always possible to identify linear combinations of $J_a,\bar J_a$ such that their OPEs do not have the term involving the structure constants --- the so-called ``no simple-pole condition'', cf.~\eqref{eq:JJ-OPE}. {In that case the correlation functions of $O$ are the same as for an $O$ constructed out of free bosons, which in turn implies that the deformation can be completed to all orders in conformal perturbation theory in the deformation parameter. For deformations satisfying only the weak CS condition on the other hand there is no guarantee that they remain marginal beyond lowest order in the deformation parameter.}
In the literature on marginal deformations of WZW models, see e.g.~\cite{Chaudhuri:1992yca,Hassan:1992gi,Kiritsis:1993ju,Tseytlin:1993hm,Forste:2003km,Israel:2004vv,Israel:2004cd,Orlando:2006cc,Detournay:2005fz,Fredenhagen:2007rx}, we did not find examples that satisfy the weak CS condition but not the strong one. Here we will construct such examples by involving sufficient components of the chiral and antichiral currents of $SL(2,\mathbb R)$. In this sense our results identify new directions to explore the conformal manifold.

In~\cite{Hassan:1992gi,Henningson:1992rn,Kiritsis:1993ju} it was argued that $O(d,d)$ transformations provides the correct language to obtain the exact (in the deformation parameter) version of the CFTs deformed by the abelian current-current operators. Indeed, such  transformations do not break the isometries involved in the deformation and one can show that the derivative of the action with respect to the deformation parameter is given by
\begin{equation}
\frac{dS}{d\eta}=-\frac{T}{2}\int d^2\sigma\,J^\eta\bar J^\eta\,,
\end{equation}
where $J^\eta$ and $\bar J^\eta$ are (anti)chiral currents of the \emph{deformed} theory corresponding to the isometries involved in the deformation. This makes it clear that the infinitesimal deformation can be integrated to a finite one.
The relevant so-called $\beta$-shifts of $O(d,d)$, corresponding to a simple shift of the $B$-field of the dual model obtained by performing T-duality on two $U(1)$ isometries, are also known as TsT (T-duality, shift, T-duality) transformations~\cite{Lunin:2005jy,Frolov:2005ty,Frolov:2005dj,Alday:2005ww}. A TsT transformation exploits an abelian $U(1)^2$ global symmetry of the sigma model to construct a deformation parameterised by a continuous deformation parameter. Since the deformation is constructed by exploiting T-duality, on-shell the deformed model is equivalent to the undeformed one, and the deformation can be equivalently understood as a twist in the boundary conditions in the compact direction of the worldsheet.  

TsT transformations are known to belong to a larger class of deformations of sigma models, usually called Yang-Baxter (YB) deformations. They first appeared in the context of integrable models, since the deformations do not break the classical integrability of the original model~\cite{Klimcik:2008eq,Delduc:2013qra,Kawaguchi:2014qwa}. YB deformations are particularly interesting in the context of the $AdS/CFT$ correspondence, since they can be used to generate string backgrounds deforming the standard ones appearing in the $AdS_{d+1}/CFT_d$ dualities. Particularly important cases are those for which integrability techniques may be applied. In the case of $AdS_5/CFT_4$ it was proposed that the deformations of the $AdS_5\times S^5$ background should correspond to non-commutative deformations of $\mathcal N=4$ super Yang-Mills~\cite{Matsumoto:2014gwa,vanTongeren:2015uha,vanTongeren:2016eeb,Araujo:2017jkb}.
The name YB comes from the fact that the deformation is controlled by an object $R$ which is an element of $\alg g\wedge \alg g$ (where $\alg g$ is the algebra of isometries of the starting background) and solves the classical Yang-Baxter equation\footnote{There is also a version of these models where $R$ solves instead the \emph{modified} CYBE~\cite{Klimcik:2008eq,Delduc:2013qra}. The story in that case is quite different and we will not consider it here.} (CYBE) on $\alg g$. The simplest solutions to the CYBE are the so-called abelian $R$-matrices, e.g. $R=T_1\wedge T_2$ with $T_i\in \alg g$ and $[T_1,T_2]=0$. In this case the CYBE is trivially satisfied because the relevant structure constants vanish. Abelian YB deformations were shown to be equivalent to TsT transformations in~\cite{Osten:2016dvf}. On compact algebras, the CYBE only admits abelian solutions. On non-compact algebras, instead, more interesting non-abelian solutions (i.e. $R$-matrices constructed out of generators of a non-abelian subalgebra) are possible. 

Originally, in the construction of the YB-deformed sigma models, the CYBE was necessary in order to preserve the classical integrability. Later it was understood that YB deformations may be obtained from non-abelian T-duality (NATD)~\cite{Hoare:2016wsk,Borsato:2016pas}. That interpretation revealed a consistent generalisation of what is known about TsT transformations, since it became clear that YB deformations correspond to a shift of the $B$-field of the dual (undeformed) sigma model; the deformed model is then obtained by applying NATD on the subalgebra of $\alg g$ where $R$ is non-degenerate. After restricting the domain, $R$ may be inverted and its inverse $R^{-1}$ is a Lie algebra 2-cocycle. The shift of the dual $B$-field is given by this 2-cocycle. In other words, it is possible to go beyond the construction related to integrable models, and understand the CYBE as being a constraint necessary to shift the dual $B$-field without modifying its field strength $H=dB$. Consistently with this interpretation, in~\cite{Lust:2018jsx,Sakamoto:2018krs} it was proposed  to identify YB deformations with the $\beta$-shifts of a larger group extending the known $O(d,d)$ group of abelian T-duality, that in~\cite{Lust:2018jsx} was dubbed ``non-abelian T-duality group''.
The  logic of NATD/$\beta$-shifts may be used to construct YB deformations of generic sigma models~\cite{Lust:2018jsx,Sakamoto:2018krs,Borsato:2018idb}, beyond those for which YB deformations were first introduced, the Principal Chiral Model and (super)cosets.\footnote{Ref.~\cite{Bakhmatov:2017joy} put forward a first proposal for a set of transformation rules to go beyond these cases.}   Here we will use the transformation rules of~\cite{Borsato:2018idb}, which  were derived from the NATD construction and have the advantage of being applicable to a generic background with isometries  (even when the initial $G-B$ is not invertible, as is the case for the background metric and Kalb-Ramond field that we have to consider in this paper).

Because of their realization via NATD, YB models will be Weyl invariant at least to one loop in $\sigma$--model perturbation theory (and exactly in the deformation parameter), provided that the Lie algebra on which the $R$-matrix is non-trivial is unimodular (i.e. the trace of its structure constants vanish). In this case the deformed background solves the standard supergravity equations of motion. When the algebra is not unimodular there is a potential Weyl anomaly \cite{Alvarez:1994np,Elitzur:1994ri}. In that case the resulting background solves instead a generalization of the standard supergravity equations \cite{Arutyunov:2015mqj,Wulff:2016tju} controlled by a Killing vector field $K$. Even in these non-unimodular cases, it can happen that there is actually no Weyl anomaly. This is reflected in the fact that the generalized supergravity equations can have ``trivial'' solutions, i.e. solutions with $K\neq0$ but where nevertheless the other fields solve the standard supergravity equations \cite{Wulff:2018aku}. In \cite{Wulff:2018aku} it was shown that this can happen if $K$ is null. In appendix \ref{app:trivial} we show that this condition can be weakened and $K$ does not have to be null if the one-form {$\rm X$ appearing in the generalized supergravity equations takes the form $\rm X=d\phi+\tilde K$ with $\phi$ the dilaton and $\tilde K$ another Killing vector}.
We will see that YB deformations of the $AdS_3\times S^3$ WZW model give rise also to ``trivial'' solutions, both ones with $K^2=0$ and with $K^2\neq0$. We will also find some examples with a genuine anomaly, corresponding to $K$ not being null (and $\tilde K$ not defining a Killing vector).

As we will discuss in more detail in section~\ref{sec:YB}, at leading order in the deformation parameter YB deformations correspond to current-current deformations. We are therefore led to study YB deformations of strings on backgrounds containing an $AdS_3$ subspace, expecting to find marginal deformations of the corresponding WZW model. Particularly  interesting for us are the deformations that do not solve the strong version of the CS condition, but only the weak one. We will construct explicitly such examples. Such possibilities are allowed because we exploit also the non-compact part of the current algebra to generate the deformations. 

The paper is organised as follows. In section~\ref{sec:WZW} we review some aspects of the $SL(2,\mathbb R)\times SU(2)$ WZW model and of marginal current-current deformations that are important for our discussion. In section~\ref{sec:YB} we review the transformation rules of YB deformations and explain in which cases we can understand them as compositions of simpler YB transformations. We will also explain the connection to the marginal current-current deformations. Using the classification of $R$-matrices in appendix~\ref{app:R-matrices}, we later study deformations of $AdS_3$ and $AdS_3\times S^3$. We give our conclusions in section~\ref{sec:disc}.
Appendix~\ref{app:field-red} collects some details on the field redefinition used in section~\ref{sec:YB}, and  appendix~\ref{app:onshell} discusses the on-shell equivalence of the YB models to the undeformed ones. In appendix~\ref{app:sl3} we consider the case of the $\alg{sl}_3$ algebra, which is separate from the rest of the paper. In appendix~\ref{app:trivial} we extend the triviality condition of~\cite{Wulff:2018aku}.

\section{Wess-Zumino-Witten model and marginal current-current deformations}\label{sec:WZW}
In this section we review certain aspects of WZW models and their marginal current-current deformations. Although the discussion can be made general, for concreteness we will take the example of the $SL(2,\mathbb R)\times SU(2)$ WZW model, since it is important for string theory applications and it already contains all the salient features.

\subsection{The $AdS_3\times S^3$ sigma model}\label{sec:sigmamodel}
We start with a sigma model describing the propagation of a string in $AdS_3\times S^3$, that can be viewed (after adding four free bosons) as the bosonic sector of the superstring . The sigma model action is\footnote{
We work with a Lorentzian worldsheet and we introduce worldsheet coordinates $\sigma^\pm=\sigma^0\pm\sigma^1$, so that $\eta^{+-}=\eta^{-+}=-2$, $\epsilon^{+-}=-\epsilon^{-+}=-2$ and $d^2 \sigma=\frac12d\sigma^+d\sigma^-$. We also use the standard notation $\sigma,\bar \sigma$ in place of $\sigma^+,\sigma^-$, as well as $\partial =\partial_\sigma,\bar \partial=\partial_{\bar \sigma}$.}
\be
S=\frac{k}{2\pi}\int d^2\sigma\ \left(\frac{-\partial x^-\bar\partial x^++\partial z\bar\partial z}{z^2}+\frac{1}{4} \partial\phi_i\bar\partial\phi_i+\frac{1}{2} \partial\phi_2\bar \partial\phi_1\sin \phi_3\right)\,.
\ee
Here we are considering the pure NSNS background, and $k$ will be the level of the WZW model. The string tension is $T=k/\pi$, and the metric and $B$-field appearing in the sigma model action follow from $S=T\int  d^2 \sigma\ L=\frac{T}{2}\int d^2 \sigma\ \partial x^m (G_{mn}-B_{mn})\bar \partial x^n$ and are\footnote{In our conventions $B=\frac{1}{2} B_{nm}dx^m\wedge dx^n$.}
\be
\begin{aligned}
ds^2&=ds_{\text{AdS}_3}^2+ds_{\text{S}^3}^2=
\frac{-dx^+dx^-+dz^2}{z^2}
+\frac{1}{4}\left[d\phi_3^2+\cos^2\phi_3d\phi_1^2+(d\phi_2+\sin \phi_3 d\phi_1)^2\right]\,,\\
B&=\frac{dx^+\wedge dx^-}{2z^2}-\frac{1}{4}\sin \phi_3 d\phi_1\wedge d\phi_2\,.
\end{aligned}
\ee
$AdS_3$ is parameterised by the boundary coordinates $x^\pm$ and the radial coordinate $z$, while the angles $\phi_i$ parameterise the sphere. The $AdS_3$ metric admits the following Killing vectors 
\be
\begin{aligned}
&k_{0}^m\partial_m=x^+\partial_{x^+}+\tfrac{1}{2}z\partial_{z}\,,\qquad
&&k_{+}^m\partial_m=\partial_{x^+}\,,\qquad
&&&k_{-}^m\partial_m=-(x^+)^2\partial_{x^+}-z^2\partial_{x^-}-x^+z\partial_{z}\,,\\
&\bar k_{0}^m\partial_m=-x^-\partial_{x^-}-\tfrac{1}{2}z\partial_{z}\,,\qquad
&&\bar k_{-}^m\partial_m=\partial_{x^-}\,,\qquad
&&&\bar k_{+}^m\partial_m=-(x^-)^2\partial_{x^-}-z^2\partial_{x^+}-x^-z\partial_{z}\,.
\end{aligned}
\ee
They satisfy $[k_a^m\partial_m,k_b^n\partial_n]=-f_{ab}^{\ c}k_c^p\partial_p$ (and similarly for $\bar k_a$), where $f_{ab}^{\ c}$ are the structure constants of the algebra of $SL(2,\mathbb R)$
\begin{equation}
[S_0,S_\pm]=\pm S_\pm\,,\qquad[S_+,S_-]=2S_0\,.
\label{eq:sl2}
\end{equation}
In these formulas and in the following we use a bar to distinguish the right copy of the algebra from the left copy.\footnote{We will interchangeably place the bar on an object or on its index, in other words $\bar k_a$ or $k_{\bar a}$ have the same meaning. For readability sometimes we will prefer the former.}
For the sphere we have two copies of $SU(2)$, whose algebra is generated by $T_a$ ($a=1,2,3$) with commutation relations $[T_a,T_b]=-\epsilon_{abc}T_c$.
We will not write explicitly all Killing vectors of $S^3$ since we will not need them. For our purposes it will be enough to use the two commuting Killing vectors
\begin{equation}
k_1=-\partial_{\phi_1}\qquad\mbox{and}\qquad\bar k_2=\partial_{\phi_2}\,.
\end{equation}

The sigma-model action is invariant under the transformations generated by the above Killing vectors, although in certain cases the $B$-field is not invariant but changes by a total derivative. Therefore in general the corresponding Noether currents are given by
\be
\mathcal J_{A,\pm}=k_A^{m}(G_{mn}\pm B_{mn})\partial_{\pm}x^n+j_{A,\pm}\,,
\label{eq:Noether}
\ee
where $j_{A,\pm}$ is defined by looking at the variation of the Lagrangian $\delta_A L=\varepsilon\partial_ij_A^i$ under the infinitesimal global transformation.
Because of our choice of gauge, in the $AdS_3$ part only $j_-^i$ and $\bar j_+^i$ are non-zero, and we also have $j_1^i=\bar j_2^i=0$. In the following we will ignore the transformations generated by $S_-,\bar S_+$, since for our discussion it will be enough to focus on the (maximal solvable) subalgebra generated by
\be\label{eq:subalgebra}
S_0,\quad S_+,\quad \bar S_0,\quad \bar S_-,\quad T_1,\quad \bar T_2\,.
\ee 
All Noether currents that we will need to consider will therefore have $j_{A,\pm}=0$.
Let us anticipate that these Noether currents are not always equal to the chiral (resp. antichiral) currents of the WZW description, which we shall denote by $J$ (resp. $\bar J$) and write explicitly in the next subsection. They agree up to ``improvement terms'' that do not spoil the current conservation, of the type $\epsilon^{ij}\partial_jc$ for some $c$. Restricting to the generators in~\eqref{eq:subalgebra}, for $AdS_3$ we have
\be\label{eq:noether-chiralAdS3}
\begin{aligned}
&\mathcal J_{0,+}=J_0-\tfrac{1}{2}\partial \log z\,,\qquad
&\mathcal J_{0,-}=+\tfrac{1}{2}\bar\partial \log z\,,\qquad
&\mathcal J_{+,+}=J_+\,,\qquad
&\mathcal J_{+,-}=0\,,\\
&\bar{\mathcal J}_{0,-}=\bar J_0+\tfrac{1}{2}\bar\partial \log z\,,\qquad
&\bar{\mathcal J}_{0,+}=-\tfrac{1}{2}\partial \log z\,,\qquad
&\bar{\mathcal J}_{-,-}=\bar J_-\,,\qquad
&\bar{\mathcal J}_{-,+}=0\,,
\end{aligned}
\ee
while for $S^3$
\be\label{eq:noether-chiralS3}
\begin{aligned}
&\mathcal J_{1,+}=J_1+\tfrac{1}{4}\partial\phi_1\,,\qquad
&\mathcal J_{1,-}=-\tfrac{1}{4}\bar\partial\phi_1\,,\\
&\bar{\mathcal J}_{2,-}=\bar J_2-\tfrac{1}{4}\bar\partial\phi_2\,,\qquad
&\bar{\mathcal J}_{2,+}=\tfrac{1}{4}\partial\phi_2\,.
\end{aligned}
\ee
This fact will later play an important role in our discussion.

\subsection{The $SL(2,\mathbb R)\times SU(2)$ WZW model}
The action of the WZW model is $S_{WZW}=S_1+k\Gamma$ where $k$ is the level and
\be
S_1[g] = \frac{k}{4 \pi} \int_{\partial\mathcal B} d^2\sigma \Tr(\partial^ig^{-1}\partial_i g )\,,\qquad
\Gamma[g] = -\frac{1}{6\pi}\int_{\mathcal B}d^3\sigma \epsilon^{ijk}\Tr(g^{-1} \partial_ig g^{-1} \partial_jg g^{-1} \partial_kg)\,.
\ee
Here $g$ is an element of a group $G$, depending on coordinates on $\mathcal B$, whose boundary $\partial\mathcal B$ is the worldsheet of the string.
In the following we will take the action\footnote{The relative minus sign is needed to get the correct sign in front of the $S^3$ metric. In the supersymmetric case it is naturally accounted for by the supertrace.} $S_{WZW}[g_{\rm a}]-S_{WZW}[g_{\rm s}]$ with
\be
g_{\rm a} = e^{x^+S_+}z^{2S_0}e^{-x^-S_-}\in SL(2,\mathbb R)\,,\qquad
g_{\rm s} = e^{\phi_1T_1}e^{\phi_3T_3}e^{\phi_2T_2}\in SU(2)\,.
\ee
We realise the generators of the algebra of $SL(2,\mathbb R)$ in terms of the Pauli matrices as $S_0=\sigma_3/2$, $S_+=(\sigma_1+i\sigma_2)/2$, $S_-=(\sigma_1-i\sigma_2)/2$, and similarly for $SU(2)$ we take $T_a=\tfrac{i}{2}\sigma_a$.
The Killing form is related to the trace in this representation as $K_{ab}=f_{ac}^{\ d}f_{bd}^{\ c}=4\Tr(S_aS_b)$, and similarly for $T_a$. We will use the bilinear form induced by the trace, rather than the Killing form, to raise and lower algebra indices.

The equations of motion for the action $S_{WZW}[g]$ imply chirality for the current $J=\partial g g^{-1}$, and equivalently antichirality for the current $\bar J=- g^{-1} \bar \partial g$, i.e. $\bar \partial J=0$, $\partial \bar J=0$.
We decompose the currents as $J=J_aS^a$ for $AdS_3$ and $J=J_aT^a$ for $S^3$, and similarly for $\bar J$, where $S^0=2S_0, S^\pm=S_\mp$ and $T^a=-2T_a$. Thanks to these definitions the component $J_a$ of the chiral current corresponds to the action of the generator $S_a$ (or $T_a$ in the case of the sphere) from the left, while the component $\bar J_a$ of the antichiral current corresponds to the action of the same generator from the right. The same holds for the corresponding Killing vectors $k_a$ and $\bar k_a$. In particular we have $k_a^m\partial_mg_{\rm a}=+S_ag_{\rm a}$, $\bar k_{a}^m\partial_mg_{\rm a}=-g_{\rm a}S_{a}$ for $AdS$ and $k_1^m\partial_mg_{\rm s}=-T_1g_{\rm s}$, $\bar k_{2}^m\partial_mg_{\rm s}=+g_{\rm s}T_{2}$ for the sphere.\footnote{The relative minus sign between $AdS_3$ and $S^3$ is again related to the fact that we are not using the supertrace in the action and in order to define the components of the currents.}  In our  parameterisation the components of the $SL(2)$ currents read 
\be
\begin{aligned}
&J_0={\phantom{-}}\frac{z \partial z-x^+\partial x^-}{z^2}\,,
&& J_-=\frac{x^+ (x^+\partial x^--2 z \partial z)}{z^2}+\partial x^+\,,
&&& J_+=-\frac{\partial x^-}{z^2}\,,
\\
&\bar J_0=-\frac{z \bar\partial z-x^-\bar\partial x^+}{z^2}\,,
&&\bar J_+=\frac{x^-(x^-\bar\partial x^+-2 z\bar \partial z)}{z^2}+\bar\partial x^-\,,
&&&\bar J_-=-\frac{\bar\partial x^+}{z^2}\,,
\end{aligned}
\ee
while for the $SU(2)$ currents we have
\be
\begin{aligned}
&J_1=\tfrac{1}{2} (-\partial \phi_1 - s_3\partial\phi_2 )\,,
&& J_2=\tfrac{1}{2} (-c_1 c_3 \partial \phi_2 - s_1\partial \phi_3 )\,,
&&& J_3=\tfrac{1}{2} (-c_1 \partial\phi_3 + c_3 s_1\partial\phi_2 )\,,
\\
& \bar J_2= \tfrac{1}{2} (+\bar\partial \phi_2 + s_3\bar\partial \phi_1 )\,, 
&&\bar J_1=\tfrac{1}{2} (+c_2 c_3 \bar\partial \phi_1 + s_2\bar\partial \phi_3 )\,,
&&&\bar J_3=\tfrac{1}{2} (+c_2 \bar\partial \phi_3 - c_3 s_2\bar\partial \phi_1 )\,,
\end{aligned}
\ee
where we use the shorthand notation $s_i=\sin\phi_i$, $c_i=\cos\phi_i$.
The (anti)chiral currents appear also when computing the Noether currents from the action $S_{WZW}=S_1+k\Gamma$. In fact, invariance of the WZW action under left transformations $g\to (1+\varepsilon_L+\ldots)g$ implies the conservation of the Noether current
$\mathscr J^i=\tfrac{1}{4} (\eta^{ij} -\epsilon^{ij})\partial_j gg^{-1}$
which is related to the chiral current as
\be\label{eq:noether-WZW}
\mathscr J=(\mathscr J_+,\mathscr J_{-})=(J,0)\,. 
\ee
Similarly, from the right transformations $g\to g(1+\varepsilon_R+\ldots)$ one finds the Noether current $\bar{\mathscr J}^i=-\tfrac{1}{4} (\eta^{ij} +\epsilon^{ij})g^{-1}\partial_j g$
related to the antichiral current as
\be\label{eq:noetherbar-WZW}
\bar{\mathscr J}=(\bar{\mathscr J}_{+},\bar{\mathscr J}_{-})=(0,\bar J)\,.
\ee
Conservation of the Noether current $\partial_i\mathscr J^i=0$ (respectively $\partial_i\bar{\mathscr J}^i=0$) implies chirality of $J$ (respectively antichirality of $\bar J$).
As we have already pointed out, in general these Noether currents are not the same as those of the sigma model description, which we denoted by $\mathcal J$.

\subsection{Marginal deformations}\label{sec:marginal}
In~\cite{Chaudhuri:1988qb} Chaudhuri and Schwartz considered two-dimensional CFTs with $J_a,\bar J_{a}$ satisfying current algebra relations\footnote{Since we are not normalising the currents with an explicit $k$ and we raise/lower indices with the bilinear form induced by the trace, as opposed to the Killing form, certain factors differ from \cite{Chaudhuri:1988qb}.}
\be\label{eq:JJ-OPE}
\begin{aligned}
J_a(\sigma)J^b(\sigma')\sim \frac{i\ \delta_a^b}{2k(\sigma-\sigma')^2}+ \frac{if_{ac}^{\ b}J^c}{2k(\sigma-\sigma')}\,,\\
\bar J_{ a}(\bar \sigma)\bar J^{ b}(\bar \sigma')\sim \frac{i\ \delta_{ a}^{ b}}{2k(\bar \sigma-\bar \sigma')^2}+ \frac{if_{ a c}^{\  b}\bar J^{ c}}{2k(\bar \sigma-\bar \sigma')}\,,
\end{aligned}
\ee
where we use $\sim$ since we are omitting regular terms and $f_{ab}^{\ c}$ are structure constants of a Lie algebra $\alg f$. 
The authors of~\cite{Chaudhuri:1988qb} were interested in exploring the space of marginal deformations induced by dimension $(1,1)$ operators of the type
\be
g\mathcal O(\sigma,\bar \sigma)= g c^{a b}J_a(\sigma)\bar J_{ b}(\bar \sigma)\,,
\ee
where $c^{ab}$ are constant coefficients. The above operator is ``integrably'' or exactly marginal (i.e. can be completed to all orders in conformal perturbation theory in $g$) if it has no anomalous dimension, and they found that {a necessary condition for this to hold is that}
\be\label{eq:CS-weak}
C^{abc}C^{def}  K_{ad} K_{be}K_{cf}
+ \bar C^{abc}\bar C^{def}  K_{ad} K_{be}K_{cf}=0\,,
\ee
where $K_{ab}$ is the Killing form and we have defined 
\be\label{eq:def-C-Cbar}
C^{abc}\equiv c^{da}c^{eb}f_{de}^{\ c}\,,\qquad
\bar C^{abc}\equiv c^{ad}c^{be} f_{de}^{\ c}\,.
\ee
We will call~\eqref{eq:CS-weak} the \emph{weak} Chaudhuri-Schwartz (CS) condition. Ref.~\cite{Chaudhuri:1988qb} considered only the case of compact algebras, meaning that the Killing form $K_{ab}$ is negative definite and can be taken to be diagonal. In this case~\eqref{eq:CS-weak} becomes a sum of squares of $C^{abc}$ and $\bar C^{abc}$, and it holds if and only if
\be\label{eq:CS-strong}
C^{abc}=0\,,\qquad\text{and}\qquad
\bar C^{abc}=0\,.
\ee
We will call~\eqref{eq:CS-strong} the \emph{strong} CS condition, because it is a stronger constraint in the case of non-compact algebras.
In~\cite{Chaudhuri:1988qb} it was also shown that the strong condition is equivalent to being able to rewrite 
\be
\mathcal O(\sigma,\bar \sigma)= \tilde c^{a b}\tilde J_a(\sigma)\tilde{\bar{J}}_{ b}(\bar \sigma)\,,
\ee
where $\tilde J_a(\sigma),\tilde{\bar{J}}_{ b}(\bar \sigma)$ are linear combinations of the original $J_a(\sigma)$, ${\bar{J}}_{ b}(\bar \sigma)$ such that 
\be
\begin{aligned}
\tilde J_a(\sigma)\tilde J^b(\sigma')\sim \frac{i\ \delta_a^b}{2k(\sigma-\sigma')^2}\,,\qquad
\tilde{\bar J}_{ a}(\bar \sigma)\tilde{\bar J}^{ b}(\bar \sigma')\sim \frac{i\ \delta_{ a}^{ b}}{2k(\bar \sigma-\bar \sigma')^2}\,,
\end{aligned}
\ee
i.e. the structure constants for this particular set of currents vanish. {The absence of a simple pole in these OPEs means that they are the same as similar ones for free bosons, which in turn means that the $\beta$-function for the deformation parameter $g$ vanishes and the deformation is exactly marginal. In other words, deformations corresponding to abelian subalgebras, which is the only possibility in the compact case, are exactly marginal.} When the Lie algebra $\alg f$ is non-compact {it is possible to find deformations that satisfy only the weak CS condition}, as we will see. {A priori they are not guaranteed to be marginal beyond lowest order, and indeed we will find both examples which are and those which are not.}

In fact a \emph{sufficient} condition on $c^{ab}$ such that the weak CS~\eqref{eq:CS-weak} holds is that the coefficients $c^{ab}$ identify two solvable subalgebras of $\alg f$ (one corresponding to $J_a$ and one to $\bar J_b$). This follows directly from Cartan's criterion for a solvable Lie algebra $\alg h$
\be
\alg h \text{ solvable } \iff \Tr(ab)=0, \qquad \forall a\in\alg h, b\in [\alg h,\alg h]\,.
\ee
If we are in such a situation then the two terms in~\eqref{eq:CS-weak} separately vanish because
\be
C^{abc}C^{def}  K_{cf}=0\,,\qquad\qquad
 \bar C^{abc}\bar C^{def}  K_{cf}=0\,.
\ee
In  the case of the $SL(2,\mathbb R)\times SU(2)$ WZW model, we may for example identify the two solvable subalgebras generated by $\{S_0,S_+,T_1\}$ and $\{\bar S_0,\bar S_-,\bar T_2\}$. Then if we call $Y_a\equiv\{J_0,J_+,J_1\}$ and $\bar Y_a=\{\bar J_0,\bar J_-,\bar J_2\}$ the list of the corresponding (anti)chiral currents, an operator
\be
\mathcal O(\sigma,\bar \sigma)=  c^{a b}Y_a(\sigma)\bar Y_{ b}(\bar \sigma)\,,
\ee
will be {marginal to lowest order} for \emph{generic} coefficients $c^{ab}$. Notice that generically $c^{ab}$ will not solve the strong CS condition \eqref{eq:CS-strong}.

All the solutions to the weak CS condition that we will generate from the CYBE on $\alg g=\alg f_{\sL}\oplus \alg f_{\sR}=\alg{sl}(2,\mathbb R)_{\sL}\oplus \alg{su}(2)_{\sL}\oplus \alg{sl}(2,\mathbb R)_{\sR}\oplus \alg{su}(2)_{\sR}$ will be of this type. Indeed to solve the CYBE it is enough to look at the subalgebra  generated by  $\{S_0,S_+,T_1,\bar S_0,\bar S_-,\bar T_2\}$. When they come from the YB construction, the coefficients $c^{ab}$ will obviously not be generic, and we will relate them to certain components of the $R$-matrix, see the discussion at the end of section~\ref{sec:YB-CS}. The YB construction has the advantage of giving a way to go beyond the infinitesimal deformation driven by $\mathcal O(\sigma,\bar \sigma)$, and gives a sigma-model action that is exact in the deformation parameter.

As we have argued, we expect the CYBE to give solutions to the weak CS condition in more generic situations. In appendix~\ref{app:sl3} we discuss a solution of the CYBE that provides coefficients $c^{ab}$ that solve the weak CS condition without identifying solvable subalgebras.

\section{Yang-Baxter and current-current deformations}\label{sec:YB}

\subsection{Yang-Baxter deformations}
We now review the transformation rules for the target space fields for YB deformations derived in~\cite{Borsato:2018idb}.
Given an initial sigma model with metric and Kalb-Ramond fields $G_{mn}$, $B_{mn}$, the background of the YB deformed model is given by
\be
\tilde G-\tilde B = (G-B)[1+\eta\Theta(G-B)]^{-1}\,,
\label{eq:GBtilde}
\ee
where for simplicity we are suppressing all spacetime indices. Here $\Theta^{mn}=k^m_AR^{AB}k^n_B$ is a tensor constructed out of the Killing vectors $k^m_A$ and of $R^{AB}$, which is a solution to the CYBE on the Lie algebra $\alg g$
\be\label{eq:CYBE}
R^{D[A}R^{|E|B}f_{DE}^{C]}=0\,.
\ee
In our case $\alg g= \alg f_{\sL}\oplus \alg f_{\sR}$ is the sum of a left and a right copy of $\alg f=\alg{sl}(2)\oplus \alg{su}(2)$, and $G,B$ were given in section~\ref{sec:sigmamodel}. 
The derivation of~\cite{Borsato:2018idb} assumes that the $B$-field is invariant under the isometries used in the deformation, i.e. the ones appearing in $\Theta$. This is ensured by picking the form of $B$ in section~\ref{sec:sigmamodel} and using only the isometries generated by (\ref{eq:subalgebra}), which is enough to generate any Yang-Baxter deformation (see appendix \ref{app:R-matrices}).

The deformation produces also a shift of the dilaton calculated from the determinant\footnote{In the supersymmetric case the determinant is replaced by the superdeterminant.}
\be
e^{-2\tilde\Phi}=e^{-2\Phi}\det[1+\eta\Theta(G-B)]\,,
\label{eq:Phitilde}
\ee
where $\Phi$ is the dilaton of the original background (in our case $\Phi=0$).
In general YB backgrounds are solutions to the  equations of generalised supergravity~\cite{Arutyunov:2015mqj,Wulff:2016tju}, so that in addition to the usual fields one may have also a vector $K$ computed as\footnote{Eq.~\eqref{eq:K} may be obtained from the formula derived in~\cite{Borsato:2018idb} (i.e. $K^m=\eta \Theta^{mn}n_n=\eta k^m_AR^{AB}n_B$ with $n_A=f_{AB}^{\ B}$) after using the identity $R^{AB}f_{AB}^{\ C}=-2f_{AB}^{\ A}R^{BC}$, which is a consequence of the CYBE. It is also easy to check that~\eqref{eq:K} agrees with $K^m=\eta \nabla_n^{(0)}\Theta^{mn}$ proposed in~\cite{Araujo:2017jkb}, where $\nabla_n^{(0)}$ is the covariant derivative of the original undeformed background. Indeed, using first the Killing equation for $k_A^m$ and then the anti-symmetry of $R$, we have $\nabla_m^{(0)}\Theta^{mn}=k_A^mR^{AB}\nabla_m^{(0)}k_B^n=R^{AB}k_A^m\partial_mk_B^n$. Knowing that Killing vectors satisfy $[k_A^m\partial_m,k_B^n\partial_n]=-f_{AB}^{\ C}k_C^p\partial_p$ we obtain again~\eqref{eq:K}.}
\be\label{eq:K}
K^m = -\frac{\eta}{2}R^{AB}f_{AB}^{\ C}k_C^m\,,
\ee
which is a Killing vector of the YB background $\nabla_{(m}K_{n)}=0$.
For such generalised supergravity solutions the role of (the derivative of) the dilaton is replaced by the vector\footnote{This expression applies in a gauge where $B$ is invariant under the isometry generated by $K$, $\mathcal L_KB=0$. Here we stick with the original notation of~\cite{Wulff:2016tju}. In~\cite{Borsato:2018idb} and~\cite{Wulff:2018aku} $X_m$ was used instead, but there is a risk of confusing it with $X_m={\rm X}_m+K_m$ of~\cite{Wulff:2016tju}.}
\be\label{eq:X}
{\rm X}_m=\partial_m\tilde\Phi -B_{mn}K^n\,.
\ee
When $K^m$ vanishes one goes back to a standard supergravity solution. From~\eqref{eq:K} the relation to the unimodularity condition of~\cite{Borsato:2016ose} is manifest. There exist also so-called ``trivial solutions'' of generalised supergravity~\cite{Wulff:2018aku}, i.e. when $K$ does not vanish but it decouples from the equations. A trivial solution is therefore both a solution of the generalised and the standard supergravity equations. Later we will encounter examples of this type.

\vspace{12pt}

{Let us comment on the fact that the YB transformations constructed in~\cite{Borsato:2018idb} were derived by assuming a group of left isometries for the sigma model. This is necessary in order to apply the NATD construction and twist the model with the corresponding Killing vectors $k_A$. The isometries that we will exploit here to deform $AdS_3\times S^3$, corresponding to the generators in~\eqref{eq:subalgebra}, belong both to the left and to the right copy of the symmetry group of the WZW model. The reason why we can apply the above rules of YB transformations is that the corresponding sigma model may be constructed as a coset on $SO(2,2)/SO(1,2)\times SO(4)/SO(3)$ with a WZ term. For example, focusing on $AdS_3$, we may relate the generators of $\alg{sl}(2,\mathbb R)$ to those of the conformal algebra as
\be
\begin{aligned}
&S_0=+\tfrac{1}{2}(D-J_{01}),\qquad
&&S_+=p_+,\qquad
&&&S_-=k_-,\\
&\bar S_0=-\tfrac{1}{2}(D+J_{01}),\qquad
&&\bar S_+=k_+,\qquad
&&&\bar S_-=p_-,
\end{aligned}
\ee
where e.g. $p_\pm=\tfrac12 (p_0\pm p_1)$.
Then we obtain the wanted sigma model action from $S=\tfrac{k}{2\pi}\int d^2 \sigma \Tr[g^{-1}\partial g(P+b)g^{-1}\bar \partial g]$ where $g=\exp(x^+p_++x^-p_-)\exp(D\log z)$, $P$ projects on the generators of the coset $p_i-k_i$ and $D$, and finally  $b(p_\pm-k_\pm)=\pm(p_\pm-k_\pm)$ produces the $B$-field. In this formulation the isometries that we want to exploit, generated by $S_0,S_+,\bar S_0,\bar S_-$, act from the left as $g\to hg$ and leave also the $B$-field invariant.
}
 
\vspace{12pt}

For later convenience, let us say at this point that it is easy to check that when an $R$-matrix is given by the sum of two $R$-matrices, the corresponding background can be understood as the composition of two successive YB transformations.
This is easily seen using the following identity valid when $\Theta=\Theta_1+\Theta_2$
\be
[1+\eta\Theta_1(G-B)]^{-1}\left[1+\eta\Theta_2(G-B)[1+\eta\Theta_1(G-B)]^{-1}\right]^{-1}=[1+\eta\Theta(G-B)]^{-1}\,,
\ee
which holds without assuming any property\footnote{Obviously we need to assume invertibility of the above operators.} for $\Theta_i$, neither antisymmetry nor CYBE.
Thanks to this formula it is straightforward to argue that the background metric and $B$-field of a YB deformation generated by $\Theta=\Theta_1+\Theta_2$ are equivalent to those coming from the composition of two successive deformations, e.g. first one generated by $\Theta_1$ and then one generated by $\Theta_2$ (or vice versa). The same holds for the transformation rule of the dilaton, and of the vector $K$, which is linear in $\Theta$ (or equivalently $R$). Obviously, the interpretation as YB deformations in the intermediate steps will be possible only if $\Theta_1,\,\Theta_2$ separately solve the CYBE, and if the isometries needed to implement the second deformation are not broken by the first one. In this case we will say that $\Theta$ is ``decomposable''. Apart from these subtleties, it will often prove useful to interpret a deformation generated by $\Theta=\Theta_1+\Theta_2$ as a composition of two transformations.

Later we will encounter examples in which $\Theta_1$ generates the undeformed background \emph{up to} a ($\eta$-dependent) field redefinition. In this case one can say that $\Theta=\Theta_1+\Theta_2$ is equivalent to the YB deformation generated by $\Theta_2$ alone \emph{only if} the field redefinition  $x^m\to x^m(x')=x'{}^m+\eta f^m(x')$ needed to trivialise $\Theta_1$ is compatible with $\Theta_2$. It is easy to convince oneself that the necessary compatibility condition is
\be\label{eq:compTheta2}
A^{-1}\Theta_2(x'{}^m+\eta f^m(x'))A^{-T}=\Theta_2(x'{}^m)\,,
\ee
where $A^m_{\ n}=\frac{\partial x^m}{\partial x^{'n}}$, and we are writing the explicit dependence of $\Theta_2$ on the coordinates.

\subsection{Relation to marginal current-current deformations}\label{sec:YB-CS}
Before discussing YB deformations of $AdS_3\times S^3$, let us make a simple observation: at leading order in the deformation parameter, the YB deformation is of the form $\mathcal J\mathcal J$, where $\mathcal J$ are the Noether currents of the sigma model. This is straightforwardly checked by expanding the sigma model action $S=\tfrac{T}{2}\int d^2 \sigma\ \partial x^m (\tilde G_{mn}-\tilde B_{mn})\bar \partial x^n$ to lowest order in the deformation parameter\footnote{Recall that we are restricting to the case when the isometries used to construct the deformation leave not just the action but also the Lagrangian invariant, so that $j_{Ai}=0$. This was the assumption also in the derivation done in~\cite{Borsato:2018idb}.}
\be\label{eq:expand-YB}
S=S_0-\eta\frac{T}{2}\int d^2\sigma\,R^{AB}\mathcal J_{A+}\mathcal J_{B-} +\mathcal{O}(\eta^2)\,.
\ee
While this is true for a generic sigma model, this observation is particularly interesting when the original sigma model is related to a WZW model. If the Noether currents $\mathcal J_{Ai}$ coincided with the chiral Noether currents of the WZW model $\mathscr J_{Ai}=\{\mathscr J_{ai},\bar{\mathscr J}_{ai}\}$, then we would automatically obtain a current-current deformation of the type $J\bar J$. In fact, from~\eqref{eq:noether-WZW} and ~\eqref{eq:noetherbar-WZW} one immediately finds that\footnote{Notice that the $R$-matrix may have also non-vanishing components with both indices in the left (or both in the right) copy of the algebra, but these will not contribute in the final expression, and the contributions coupling currents with the same chirality cancel out.} $\int d^2\sigma \ \epsilon^{ij}\, R^{BA}\mathscr J_{Ai} \mathscr J_{Bj}=4\int d^2\sigma\ R^{a\bar b} J_a \bar J_{\bar b}$. 
As we have seen, though, in general $\mathcal J_{A}^i= \mathscr J_{A}^i+\epsilon^{ij}\partial_jc_A$, and the  discussion is more subtle because~\eqref{eq:expand-YB} will contain additional terms together with the wanted $J\bar J$ ones. We are about to show that for YB deformations of the $AdS_3\times S^3$ sigma model these additional terms can be removed by proper field redefinitions. We can therefore relate YB deformations to the deformations of the type $c^{a b}J_a\bar J_{ b}$ considered by CS. From this discussion it is also clear that we should identify the coefficients $c^{a b}$ of CS with an ``off-diagonal'' block of the $R$-matrix. More explicitly, since $R\in \alg g\wedge \alg g$ we are solving the CYBE on an algebra which is the sum of a left and a right copy $\alg g=\alg f_{\sL}\oplus \alg f_{\sR}$, the $R$-matrix can be decomposed as
\be
R=\left(
\begin{array}{cc}
R_{\sL\sL} & R_{\sL\sR}\\
R_{\sR\sL} & R_{\sR\sR}
\end{array}
\right)
\ee
with $R_{\sL\sL}^{T}=-R_{\sL\sL}, R_{\sR\sR}^{T}=-R_{\sR\sR}$ and $R_{\sL\sR}^T=-R_{\sR\sL}$. The relation to the coefficients of CS is therefore $c^{a b}=R_{\sL\sR}^{ab}$.
We can therefore generate solutions to the (weak) CS condition from solutions of the CYBE, and we will find several non-trivial examples in the following.
Obviously abelian $R$-matrices ($R=a\wedge b, [a,b]=0$) will give solutions of the strong CS condition. When dealing with non-compact algebras the CYBE allows also for solutions that are not of the abelian type. Some of them will give coefficients $c^{ab}$ that do not solve the strong CS condition. They all solve the weak CS condition as already explained at the end of section~\ref{sec:marginal}. 

It would be very interesting to understand more deeply the relation between the space of solutions of the CYBE~\eqref{eq:CYBE} and the weak CS condition~\eqref{eq:CS-weak} in the case of a generic algebra $\alg g=\alg f_{\sL}\oplus \alg f_{\sR}$. {It is interesting to notice that in order to solve the CYBE one may need also components of the diagonal blocks $R_{\sL\sL}$ and $R_{\sR\sR}$, while in the weak CS condition these will not enter. In fact, given $R_{\sL\sR}^{a\bar b}=c^{a \bar b}$ and taking the CYBE on mixed left/right indices (where we use an explicit bar for indices of the right copy of the algebra), one gets for example $c^{d\bar a}R^{eb}f_{de}^{\ c}+c^{b\bar d}c^{c\bar e}f_{\bar d\bar e}^{\ \bar a}+R^{dc}c^{e\bar a}f_{de}^b=0$. Depending on the coefficients $c^{a\bar b}$ one may also need non-vanishing left-left $R^{ab}$ components in order to solve this equation,\footnote{Such an example is given by the fourth $R$-matrix in table~\ref{tab:non-ab-sl2}.} but the CS condition is not sensitive to them.}

Let us also comment that, differently from what was claimed in~\cite{Araujo:2018rho}, the strong CS condition~\eqref{eq:CS-strong} is \emph{not} the CYBE, not even when one further imposes the unimodularity condition (and in fact our $\mathbf R_9$ in table~\ref{tab:non-ab-sl2-su2-r4} is a counter example to that claim).

\subsection{Field redefinition}
In order to display the current-current structure of the deformed model it is convenient to write the Lagrangian in the form
\begin{equation}
L=L_0-\tfrac12\eta\mathcal J_{A+}[(1+\eta RM)^{-1}R]^{AB}\mathcal J_{B-}\,,
\label{eq:Lprime}
\end{equation}
where $L_0$ is the undeformed Lagrangian and $M_{AB}=k_A^mk_B^n(G-B)_{mn}$. This follows directly from the form of $\tilde G-\tilde B$ in (\ref{eq:GBtilde}) and the definition of the Noether currents in (\ref{eq:Noether}) upon recalling that we will pick $R$ so that the last term in $\mathcal J$ does not contribute. To compare this to the discussion of current-current deformations of the WZW model we need to perform a field redefinition that replaces the Noether currents in the above expression with the chiral currents.\footnote{It is actually enough to look at the lowest order in $\eta$ if we are considering infinitesimal deformations. {The action written in terms of the Noether currents as in~\eqref{eq:Lprime} was given also in~\cite{Araujo:2018rho} but the rewriting in terms of the chiral currents is missing there.}} In appendix~\ref{app:field-red} we find such a field redefinition for a general deformation specifying to $AdS_3\times S^3$ for concreteness.
Here we will just say that for all deformations that we consider we find that the Lagrangian can be written in the form
\begin{equation}
L=
L_0
-\tfrac12\eta\hat J_a[(1+\eta RM')^{-1}R]^{a\bar a}\hat{\bar J}_{\bar a}\,,
\label{eq:LdefAdS3}
\end{equation}
where $M'$ is a shorthand for $M$ after the field redefinition. $\hat J_a,\hat{\bar J}_{\bar a}$ are modifications of the chiral currents of the undeformed WZW model. Their explicit form is given in~\eqref{eq:JhatAdS3} for deformations of $AdS_3$, and in section~\ref{sec:def-AdS3S3} for deformations of $AdS_3\times S^3$.

At leading order in $\eta$ the above Lagrangian becomes
\begin{equation}
L=
L_0
-\tfrac12\eta J_aR^{a\bar a}\bar J_{\bar a}+\mathcal{O}(\eta^2)\,,
\end{equation}
so that the comparison to the current-current deformations considered by CS is now manifest.

\subsection{YB deformations of $AdS_3$}\label{sec:def-AdS3}
Let us start by looking at YB deformations that deform only $AdS_3$. We will start with the simplest ones which are TsT-transformations. They come from the abelian $R$-matrices of $\mathfrak{sl}(2,\mathbbm R)_{\sL}\oplus \mathfrak{sl}(2,\mathbbm R)_{\sR}$ which (up to automorphisms) are\footnote{From now on we will use the components $R^{AB}$ to construct $R=R^{AB}\mathbf T_A\wedge \mathbf T_B\in \alg g\wedge \alg g$.
} \footnote{We could consider also $R=S_0\wedge \bar S_-$, but it is related to $R=S_+\wedge \bar S_0$ if we also exchange the left and right copy of the algebra.}
\begin{equation}
S_+\wedge\bar S_-\,,\qquad S_0\wedge\bar S_0\,,\qquad S_+\wedge\bar S_0\,.
\end{equation}
For the first one we obtain, from (\ref{eq:GBtilde}) and (\ref{eq:Phitilde}), the supergravity background of a deformation of $AdS_3\times S^3\times T^4$
\be
\begin{aligned}
ds^2&=
-\frac{dx^+dx^-}{z^2-\eta}
+\frac{dz^2}{z^2}+ds_{S^3}^2+ds_{T^4}^2\,,\\
B&=\frac{dx^+\wedge dx^-}{2(z^2-\eta)}
-\frac{1}{4}\sin \phi_3 d\phi_1\wedge d\phi_2\,,\qquad\quad
e^{-2\tilde\Phi}=1-\frac{\eta}{z^2}\,.
\end{aligned}
\ee
In this case the isometries involved in the deformation procedure correspond to Noether currents that agree with the (anti)chiral currents, see~\eqref{eq:noether-chiralAdS3}. We therefore automatically get that to leading order the deformation of the Lagrangian is given by the marginal operator $\eta J_+\bar J_-$. In~\cite{Giveon:2017nie} the above  background was argued to be the dual of the ``single-trace'' $T\bar T$ deformation of the symmetric product orbifold CFT$_2$. This is also is accordance with the fact this particular YB deformation is just a TsT transformation involving the two boundary coordinates.\footnote{We can have a TsT interpretation because we can implement the sequence T-duality, shift, T-duality in terms of the coordinates $x^0,x^1$, instead of the null coordinates $x^\pm$.}
At finite order in the deformation parameter the Lagrangian is given by (\ref{eq:LdefAdS3}), (\ref{eq:JhatAdS3}) and (\ref{eq:M})
\be
L=L_0-\tfrac12\frac{\eta}{1-\frac{\eta}{z^2}}J_+\bar J_-\,.
\label{eq:TsT1}
\ee
The derivative of the action with respect to the deformation parameter is given by
\be
\frac{dS}{d\eta}=-\frac{T}{2}\int d^2\sigma \ J_+^\eta\bar J_-^\eta\,,
\ee
where $J_+^\eta=(1-\eta z^{-2})^{-1}J_+$ and $\bar J_-^\eta=(1-\eta z^{-2})^{-1}\bar J_-$ are (anti)chiral currents of the \emph{deformed} model.{ This background was analysed in~\cite{Forste:1994wp}.}
Let us also note that in this case the deformation parameter can be absorbed by a rescaling of the coordinates. There are therefore only three cases: $\eta>0$, $\eta=0$ and $\eta<0$. The first of these is not globally well behaved since the dilaton becomes imaginary when crossing $z=\sqrt\eta$. For $\eta<0$ the solution interpolates between two CFTs: the $SL(2,\mathbbm R)$ WZW model and a linear dilaton background (plus two decoupled bosons). It would be interesting to study further the implications of the existence of this interpolating solution but we will not do so here. See also~\cite{Giveon:2017nie} for a discussion on the different interpretations depending on the sign of the deformation parameter.

For the remaining two abelian $R$-matrices we obtain by a similar calculation the Lagrangians\footnote{{The $J_0\bar J_0$-deformation was considered also in~\cite{Israel:2003ry}.}}
\be
\begin{aligned}
&L=L_0-\tfrac12\eta\frac{(4+\eta)z^2}{4z^2+\eta(4+\eta)x^+x^-}J_0 \bar J_0\,,\\
&L=L_0-\tfrac12\eta\frac{z^2}{z^2+\eta  x^-}J_+ \bar J_0\,.
\end{aligned}
\ee
For all the abelian examples one finds, as already mentioned, that
\be
\frac{dS}{d\eta}=-\frac{T}{2}\int d^2\sigma \ R^{a\bar a}J_a^\eta\bar J_{\bar a}^\eta\,,
\ee
where $J_a^\eta,\bar J_{\bar a}^\eta$ are (anti)chiral currents of the deformed model. For the last two abelian examples they are
\be
\begin{aligned}
&R=S_0\wedge \bar S_0: && 
J_0^\eta=\alpha J_0\,, \qquad 
\bar J_0^\eta=\alpha \bar J_0\,,\qquad
\alpha\equiv\frac{4 z^2\sqrt{1+\frac{\eta }{2}}}{4z^2+\eta(4+\eta)x^+x^-}\,,\\
&R=S_+\wedge \bar S_0: && J_+^\eta=\alpha J_+\,, \qquad 
\bar J_0^\eta=\alpha \bar J_0\,,\qquad
\alpha\equiv\frac{z^2}{z^2+\eta  x^-}\,.
\end{aligned}
\ee
Following~\cite{Forste:1994wp}, the above result is another way to see that the YB model provides a deformation that is marginal exactly in the deformation parameter.

\vspace{12pt}

In order to find deformations that at least potentially are not TsT, one should look at the class of non-abelian $R$-matrices. The full list of non-abelian $R$-matrices of $\alg{sl}(2,\mathbb{R})_{\sL}\oplus\alg{sl}(2,\mathbb{R})_{\sR}$ (up to $SL(2,\mathbb{R})_{\sL}\times SL(2,\mathbb{R})_{\sR}$ automorphisms) is given in table~\ref{tab:non-ab-sl2}. They are special cases of the $R$-matrices for $\mathfrak{sl}(2,\mathbbm R)_{\sL}\oplus \mathfrak{sl}(2,\mathbbm R)_{\sR}\oplus\mathfrak{su}(2)_{\sL}\oplus \mathfrak{su}(2)_{\sR}$ classified in appendix \ref{app:R-matrices}.
\begin{table}
\begin{center}
\begin{tabular}{|l|c|c|c|}
\hline
& & & \\
$R=R^{AB}\mathbf T_A\wedge \mathbf T_B$ & Deformation?& $c^{ab}J_a\bar J_b$ & Strong CS?\\
\hline
$R_1=S_0\wedge S_+$ & Trivial deformation & 0& Yes\\
$R_2=(S_0\mp\bar S_-)\wedge S_+$ & TsT & $\pm J_+\bar J_-$& Yes\\
$R_3=(S_0-a\bar S_0)\wedge S_+$ & TsT & $aJ_+\bar J_0$& Yes\\
$R_4=(S_0-\bar S_0)\wedge (S_+\pm\bar S_-)$ & Not SUGRA & $J_+\bar J_0\pm J_0\bar J_-$& No\\
$R_5=S_0\wedge S_+\pm\bar S_0\wedge \bar S_-+\lambda {S_+\wedge \bar S_-}$ & {TsT} & $\lambda {J_+\bar J_-}$& Yes\\
\hline
\end{tabular}
\caption{Non-abelian $R$-matrices of $\alg{sl}(2)_{\sL}\oplus\alg{sl}(2)_{\sR}$ up to $SL(2,\mathbb{R})_{\sL}\times SL(2,\mathbb{R})_{\sR}$ inner automorphisms and swaps of $\sL\leftrightarrow \sR$. For convenience we also write the marginal operators that they give rise to, and whether they satisfy the strong CS condition.}
\label{tab:non-ab-sl2}
\end{center}
\end{table}
Analysing the first example $R_1=S_0\wedge S_+$ one finds that, after the field redefinition $x^+\to x^+-\frac{\eta}{2}\log z$, not only the leading order in $\eta$ vanishes in the action, but also all the higher orders. This is obvious from eq. (\ref{eq:LdefAdS3}) and the fact that $R$ with an anti-chiral index vanishes. In other words the deformation is trivial, since its effect is to give the undeformed $AdS_3$ background in new ($\eta$-dependent) coordinates. To be more precise, from $R_1$ we get back undeformed $AdS_3$ \emph{up to} a non-vanishing $K=-\eta\partial_{x^+}$, from (\ref{eq:K}). This is of course a ``trivial solution''  of generalised supergravity (i.e. one that solves also standard supergravity upon dropping $K$, notice  that $K$ is null).\footnote{Obviously a similar discussion holds also for $R=\bar S_0\wedge \bar S_-$.}

The above result for $R_1=S_0\wedge S_+$ turns out to be useful to analyse some of the following examples in the table. 
It is easy to see that $R_2$ and $R_3$ in table~\ref{tab:non-ab-sl2} are of the form $R= S_0\wedge S_++R'$ and that in both cases $R'$ is compatible with the field redefinition  that trivialises the effect of the piece $S_0\wedge S_+$, see~\eqref{eq:compTheta2}. Therefore these two $R$-matrices are equivalent to two of the TsT transformations already discussed,\footnote{Also in this case the equivalence holds up to a non-vanishing $K=-\eta\partial_{x^+}$ that is null and decouples from the equations of generalised supergravity.} generated by $S_+\wedge \bar S_-$ and $S_+\wedge \bar S_0$. Alternatively it is easy to see this directly from the Lagrangian in (\ref{eq:LdefAdS3}) and (\ref{eq:JhatAdS3}).

The last two $R$-matrices in table~\ref{tab:non-ab-sl2}, instead, give backgrounds that are not of this type. From (\ref{eq:K}) we find that they have $K=-\eta(\partial_{x^+}\pm\partial_{x^-})$ and $K=-\eta(\partial_{x^+}\mp\partial_{x^-})$ respectively, neither of which is null. The only way they can give solutions of standard supergravity is then, {as shown in appendix \ref{app:trivial}, if ${\rm X}=d\phi+\tilde K$ with $\phi$ the dilaton and $\tilde K$ an independent Killing vector field. We can extract $\tilde K$ from the equation $dK+i_{\tilde K}H=0$ in (\ref{eq:trivial}) and for $R_4$ one finds
$\tilde K=\eta(\partial_{x^+}\mp\partial_{x^-})\pm\eta^2z^{-1}\partial_z+\mathcal O(\eta^3)$ which, at lowest order, differs from $K$ only in the sign of the $\partial_{x^+}$-term. However from the lowest order correction to the action $J_+\bar J_0\pm J_0\bar J_-$ we read off the deformed metric 
\begin{equation}
z^{-2}(-dx^-dx^++dzdz)-\eta z^{-4}(zdz(dx^-\mp dx^+)+(\pm x^+-x^-)dx^+dx^-)+\mathcal O(\eta^2)\,,
\end{equation}
which is only invariant under $\delta x^+=\eta\epsilon$, $\delta x^-=\pm\eta\epsilon$ which is generated by $K$, and not under $\delta x^+=\eta\epsilon$, $\delta x^-=\mp\eta\epsilon$, $\delta z=\pm\eta^2\epsilon z^{-1}$ which is generated by $\tilde K$. Therefore $\tilde K$ is not Killing and the $R_4$-background is not a solution to standard supergravity. Note that it only fails to be a solution at order $\eta^2$ which is consistent with the fact that the leading-order deformation  satisfies the weak CS condition.}
 We will not consider this background further here since our interest is mainly in string theory applications. For $R_5$, instead, one can show that $\tilde K$ is a Killing vector of the deformed metric and it satisfies~\eqref{eq:trivial}, meaning that we get a ``trivial solution'' of generalised supergravity for generic $\lambda$. Actually, using (\ref{eq:LdefAdS3}) and (\ref{eq:JhatAdS3}), for $R_5$ we find the Lagrangian\footnote{For concreteness we take $R_5=S_0\wedge S_++\bar S_0\wedge \bar S_-+\lambda {S_+\wedge \bar S_-}$, but similar results apply also for $R_5=S_0\wedge S_+-\bar S_0\wedge \bar S_-+\lambda {S_+\wedge \bar S_-}$.}
\begin{equation}
L=L_0+{\frac{\eta  z^2 (\eta -4 \lambda )}{2 \left(\eta  (\eta -4 \lambda )+4 z^2\right)}}J_+\bar J_-\,,
\label{eq:lambda0}
%
\end{equation}
which is exactly that of the TsT example in~\eqref{eq:TsT1} with {$\eta\rightarrow \eta\lambda-\eta^2/4$}. This background therefore provides an example of the more general kind of trivial solution of the generalized supergravity equations discussed in appendix \ref{app:trivial} for which $K^2\neq0$.

{This example also shows how the identification of the deformation parameters can be non-trivial. In fact, although at leading order the marginal deformation is given only by $\eta\lambda J_+\bar J_-$, the deformation exact in $\eta$ shows that the deformation parameter of the TsT is instead $\eta\lambda-\eta^2/4$, and in particular it does not vanish even when $\lambda=0$.}

It is interesting to look at the marginal operators of the type $c^{ab}J_a\bar J_b$  that are generated by each $R$-matrix. We write them for convenience in table~\ref{tab:non-ab-sl2}. While they all solve the weak CS condition, they all solve also the strong CS condition {(which guarantees that they are exactly marginal)} except for the fourth one {(which we have seen fails to be marginal beyond lowest order)}. 

\subsection{YB deformations of $AdS_3\times S^3$}\label{sec:def-AdS3S3}
Looking at deformations of $AdS_3\times S^3$ gives a richer set of possibilities. In this case we want to solve the CYBE on the algebra $\alg{sl}(2)_{\sL}\oplus\alg{sl}(2)_{\sR}\oplus\alg{su}(2)_{\sL}\oplus\alg{su}(2)_{\sR}$.

The simplest example to start with is $R=S_+\wedge \bar T_2$. This is an abelian $R$-matrix and therefore corresponds to a TsT mixing $AdS_3$ and $S^3$. In~\cite{Chakraborty:2018vja,Apolo:2018qpq} it was argued that this background is dual to the single-trace $T\bar J$ deformation of the CFT$_2$. From the YB procedure one explicitly finds
\be
\begin{aligned}
ds^2&=ds_{\text{AdS}_3}^2+ds_{\text{S}^3}^2+\frac{\eta}{4z^2}dx^-(d\phi_2+2\sin \phi_3d\phi_1),\\
B&=B_0
+\frac{\eta  dx^-\wedge(d\phi_2 +2 \sin \phi_3 d\phi_1)}{8 z^2},
\qquad e^{-2\Phi}=1.
\end{aligned}
\ee
From~\eqref{eq:noether-chiralS3} we see that the Noether current $\bar{\mathcal J}_2$ differs from the antichiral current $\bar J_2$, and therefore field redefinitions are needed in order to put the action in the form that makes the chirality structure manifest. After the field redefinition $x^+\to x^+-\frac{\eta}{4}\phi_2$, or equivalently by looking directly at (\ref{eq:Ldef}), (\ref{eq:Jhat}) and (\ref{eq:Jbarhat}) the Lagrangian becomes
\be
L=L_0-\tfrac12\eta J_+\bar J_2\,.
\ee
This deformation is special since the leading linear order is exact. Obviously $dS/d\eta=- \tfrac{T}{2} \int J_+\bar J_{2}$.

We will now focus on non-abelian $R$-matrices. These are classified for $\alg{sl}(2)_{\sL}\oplus\alg{sl}(2)_{\sR}\oplus\alg{su}(2)_{\sL}\oplus\alg{su}(2)_{\sR}$ in appendix~\ref{app:R-matrices}. Since the CYBE on the two copies of $\alg{su}(2)$ implies that the subset of generators from this part of the algebra should be abelian, our classification is useful also to study deformations of the full $AdS_3\times S^3\times T^4$ or the more generic $AdS_3\times S^3\times S^3\times S^1$. Every time an $\alg{su}(2)$ generator appears, it may as well be replaced by another compact generator, as long as all compact generators involved form an abelian subalgebra. For concreteness we will only look at deformations of $AdS_3\times S^3$. The rank-2 non-abelian $R$-matrices are collected in table~\ref{tab:non-ab-sl2-su2-r2}. Since they are special cases of the rank-4 $R$-matrices, listed in table~\ref{tab:non-ab-sl2-su2-r4}, we will not consider them separately.

\begin{table}[ht]
\begin{center}
\begin{tabular}{|l|c|}
\hline
&   \\
$R=R^{AB}\mathbf T_A\wedge \mathbf T_B$ & Deformation?\\
\hline
&   \\
$\mathbf R_1=(S_0-\bar S_0+T)\wedge (S_+\pm\bar S_-)$ & $R_4$+TsT$\implies$ not SUGRA  \\
$\mathbf R_2=(S_0-a\bar S_0+T)\wedge S_+$ & $R_3$+TsT$\implies$ TsT \\
$\mathbf R_3=(S_0\mp\bar S_-+T)\wedge S_+$ & $R_2$+TsT$\implies$ TsT  \\
\hline
\end{tabular}
\begin{tabular}{|c|c|c|}
\hline
& &  \\
$R$& $c^{ab}J_a\bar J_b$ & Strong CS?\\
\hline
&  & \\
$\mathbf R_1$ &  $\pm(J_0+aJ_1)\bar J_-+J_+(\bar J_0+b\bar J_2)$ & No \\
$\mathbf R_2$ & $J_+(a\bar J_0+b\bar J_2)$& Yes \\
$\mathbf R_3$ & $J_+(\pm\bar J_-+b\bar J_2)$& Yes \\
\hline
\end{tabular}
\caption{Rank-2 non-abelian $R$-matrices of $\alg{sl}(2)_{\sL}\oplus\alg{sl}(2)_{\sR}\oplus \alg{su}(2)_{\sL}\oplus\alg{su}(2)_{\sR}$ up to $SL(2,\mathbb{R})_{\sL}\times SL(2,\mathbb{R})_{\sR}\times SU(2)_{\sL}\times SU(2)_{\sR}$ inner automorphisms and swaps of $\sL\leftrightarrow \sR$. With $T$ we denote a generic linear combination of $T_1, \bar T_2$. }
\label{tab:non-ab-sl2-su2-r2}
\end{center}
\end{table}

\begin{table}
\begin{center}
\begin{tabular}{|l|c|}
\hline
& \\
$R=R^{AB}\mathbf T_A\wedge \mathbf T_B$ & Deformation?\\
\hline
&  \\
$\mathbf R_4=\mathbf R_1+a T_1\wedge \bar T_2$ & $\mathbf R_1+$TsT$\implies$not SUGRA \\
$\mathbf R_5=\mathbf R_2+b T_1\wedge \bar T_2$ & $\mathbf R_2+$TsT$\implies$TsT  \\
$\mathbf R_6=(S_0+T)\wedge S_++T' \wedge (\bar S_0+T'')$ & SUGRA  \\
$\mathbf R_7=(S_0+T)\wedge S_++T' \wedge (\bar S_-+T'')$ & TsT  \\
$\mathbf R_8=\mathbf R_3+a T_1\wedge \bar T_2$ & $\mathbf R_3+$TsT$\implies$TsT \\
$\mathbf R_9=(S_0+\bar S_0+T)\wedge T'+S_+\wedge \bar S_-$ & SUGRA, not TsT \\
$\mathbf R_{10}={(S_0+T)\wedge S_+\pm(\bar S_0+T')\wedge \bar S_-+\lambda S_+\wedge \bar S_-}$ & {TsT}  \\
\hline
\end{tabular}
\begin{tabular}{|c|c|c|}
\hline
&  & \\
$R$& $c^{ab}J_a\bar J_b$ & Strong CS?\\
\hline
& &  \\
$\mathbf R_4$  & $(J_0+aJ_1)\bar J_-+J_+(\bar J_0+b\bar J_2)+a J_1\bar J_2$ & No \\
$\mathbf R_5$  & $J_+(a\bar J_0+b\bar J_2)+bJ_1\bar J_2$ & Yes \\
$\mathbf R_6$ & $aJ_+\bar J_2+bJ_1\bar J_0+cJ_1\bar J_2$ & Yes \\
$\mathbf R_7$  &  $aJ_+\bar J_2+bJ_1\bar J_-+cJ_1\bar J_2$ & Yes \\
$\mathbf R_8$ & $J_+(\bar J_-+b\bar J_2)+aJ_1\bar J_2$ & Yes \\
$\mathbf R_9$ & $aJ_0\bar J_2+bJ_1\bar J_0+cJ_1\bar J_2+J_+\bar J_-$ & No ($a\neq 0$ or $b\neq 0$)\\
$\mathbf R_{10}$ & ${cJ_+\bar J_2+d J_1\bar J_-+\lambda J_+\bar J_-}$ & {Yes} \\
\hline
\end{tabular}
\caption{Rank-4 non-abelian $R$-matrices of $\alg{sl}(2)_{\sL}\oplus\alg{sl}(2)_{\sR}\oplus \alg{su}(2)_{\sL}\oplus\alg{su}(2)_{\sR}$ up to $SL(2,\mathbb{R})_{\sL}\times SL(2,\mathbb{R})_{\sR}\times SU(2)_{\sL}\times SU(2)_{\sR}$ inner automorphisms and swaps of $\sL\leftrightarrow \sR$. With $T,T',T''$ we denote  generic linear combinations of $T_1, \bar T_2$.}
\label{tab:non-ab-sl2-su2-r4}
\end{center}
\end{table}

To compute the Killing vector $K$ that appears in the generalized supergravity equations we note that eq. (\ref{eq:K}) implies that only the non-abelian generators matter and we can set $T=T'=T''=0$ (where these are   generic linear combinations of $T_1, \bar T_2$). We find therefore the same answer as for the $SL(2,\mathbbm R)$ case in the previous section and comparing to that analysis we see that only $\mathbf R_4$ (and therefore $\mathbf R_1$) does not give a supergravity solution. Since we are interested in string theory applications we will focus on the analysis of $\mathbf R_5$ through $\mathbf R_{10}$. Recall that in this way we automatically consider also the rank-2 $R$-matrices $\mathbf R_2$ and $\mathbf R_3$.  From (\ref{eq:Ldef}), (\ref{eq:Jhat}) and (\ref{eq:Jbarhat}) it follows that (after the field redefinitions) the Lagrangian takes the form
\begin{equation}
L=
L_0
-\tfrac12\eta\hat J_a[(1+\eta RM')^{-1}R]^{a\bar a}\hat{\bar J}_{\bar a}\,,
\end{equation}
where for $\mathbf R_5$--$\mathbf R_8$
\begin{equation}
\hat J_a=J_a+\eta\delta_a^0y_iY^i_AR^{A+}J_+\,,\qquad
\hat{\bar J}_{\bar a}=\bar J_{\bar a}-\eta\delta_{\bar a}^{\bar0}Y^i_AR^{A\bar-}y_i\bar J_-\,,
%
\end{equation}
while for $\mathbf R_9$
\begin{equation}
\hat J_a=e^{\delta_a}J_a\,,\qquad
\hat{\bar J}_{\bar a}=e^{\bar\delta_{\bar a}}\bar J_{\bar a}\,,\qquad
\delta_+=-\bar\delta_{\bar-}=-\eta Y^i_AR^{A0}y_i\,.
%
\end{equation}
It is not hard to show that $\mathbf R_5$, $\mathbf R_7$ and $\mathbf R_8$ are in fact equivalent to TsT transformations. Consider the first one. The deformed Lagrangian is
\begin{equation}
L=
L_0
-\tfrac12\eta[J_a+\eta\delta_a^0y_iY^i_AR^{A+}J_+][(1+\eta RM')^{-1}R]^{a\bar a}\bar J_{\bar a}\,.
\end{equation}
However, due to the form of the $R$-matrix and the fact that $M'_{+A}=0$ this is equal to
\begin{equation}
L=
L_0
-\tfrac12\eta J_a[(1+\eta RM')^{-1}R]^{a\bar a}\bar J_{\bar a}\,,
\end{equation}
where furthermore we can replace $R$ by the abelian $R$-matrix obtained by dropping the $S_0\wedge S_+$-term in $\mathbf R_5$. This deformation therefore reduces to a sequence of commuting TsT transformations. The same conclusion applies to $\mathbf R_8$ as is easily seen from the form of the $R$ matrix. For $\mathbf R_7$ we have
\begin{equation}
L=
L_0
-\tfrac12\eta[J_a+\eta\delta_a^0y_iY^i_AR^{A+}J_+][(1+\eta RM')^{-1}R]^{a\bar a}[\bar J_{\bar a}-\eta\delta_{\bar a}^{\bar0}Y^i_AR^{A\bar-}y_i\bar J_-]\,,
\end{equation}
but again the terms with $a=0$ and $\bar a=\bar0$ drop out due to the form of the $R$-matrix and the fact that $M_{+A}=0$ and this reduces to
\begin{equation}
L=
L_0
-\tfrac12\eta J_a[(1+\eta RM')^{-1}R]^{a\bar a}\bar J_{\bar a}\,.
\end{equation}
It is also not hard to see, using again the form of the matrices $M$ and $R$, that one can again replace $R$ by the abelian $R$-matrix obtained by dropping the $S_0\wedge S_+$-term in $\mathbf R_7$. The resulting background is therefore also a TsT. 
The fact that $\mathbf R_5$, $\mathbf R_7$ and $\mathbf R_8$ are equivalent to TsT backgrounds may be argued also from the fact that the $R$-matrices are abelian up to the $S_0\wedge S_+$-term, and that~\eqref{eq:compTheta2} holds.

For $\mathbf R_6$ the deformed Lagrangian is
\begin{equation}
L=
L_0
-\tfrac12\eta[J_a+\eta\delta_a^0y_iY^i_AR^{A+}J_+][(1+\eta RM')^{-1}R]^{a\bar a}\bar J_{\bar a}\,.
\end{equation}
Again, using the form of $R$ and the fact that $M_{+A}=0$, this simplifies to
\begin{equation}
L=
L_0
-\tfrac12\eta J_a[(1+\eta RM')^{-1}R]^{a\bar a}\bar J_{\bar a}\,.
\end{equation}
However, in this case we cannot get an equivalent background simply by dropping the $S_0\wedge S_+$-term in $\mathbf R_6$.\footnote{One way to see this is that $\mathbf R_6-R_1$ is not compatible with the coordinate redefinition needed to undo the transformation with $R_1=S_0\wedge S_+$ (see eq.~\eqref{eq:compTheta2}), therefore one does not expect to be able to undo the effect of the $R_1$ piece of the $R$-matrix.} Let us work out the background for the simplest case, $T=T''=0$ and $T'=aT_1+b\bar T_2$. One finds
\begin{equation}
L
=
L_0
-\eta f^{-1}
\Big(
2b\eta J_+\bar J_2
+2ab\eta(1-\eta\frac{x^-}{z^2})J_1\bar J_2
-\tfrac12\eta^2(a^2+b^2-2ab\sin\phi_3)J_+\bar J_0
+8aJ_1\bar J_0
\Big)
%
%
%
%
%
%
\end{equation}
where $f=16+\eta^2(a^2+b^2-2ab\sin\phi_3)[1-\eta x^-/z^2]$. One can show that in fact the (anti)chiral currents entering this action $J_+$, $J_1$, $\bar J_0$ and $\bar J_2$ extend to chiral currents to all orders in $\eta$, i.e. the corresponding isometries are not broken by the deformation. Since this is a characteristic feature of TsT backgrounds it is natural to guess that this background can be generated in that way. It is not hard to show this explicitly in the special cases $a=1$, $b=0$ and $a=0$, $b=1$ in which cases it is equivalent to the backgrounds generated by
\begin{equation}
R=T_1\wedge\bar S_0-\frac{\eta^2}{16}S_+\wedge\bar S_0\qquad\mbox{and}\qquad
R={\frac{4\eta}{16+\eta^2}S_+\wedge\bar T_2}-\frac{\eta^2}{16+\eta^2}S_+\wedge\bar S_0\,,
\end{equation}
respectively.

$\mathbf R_9$ is the only unimodular example, i.e. the only one with $K=0$. The deformed Lagrangian is
\begin{equation}
L=
L_0
-\tfrac12\eta\hat J_a[(1+\eta RM')^{-1}R]^{a\bar a}\hat{\bar J}_{\bar a}\,,
\end{equation}
with
\begin{equation}
\hat J_a=e^{\delta_a}J_a\,,\qquad
\hat{\bar J}_{\bar a}=e^{\bar\delta_{\bar a}}\bar J_{\bar a}\,,\qquad
\delta_+=-\bar\delta_{\bar-}=-\eta Y^i_AR^{A0}y_i
%
\end{equation}
For simplicity we will set $T=0$ and $T'=T_1$. We can argue that this example is not equivalent to a TsT as follows. To order $\eta^2$ the Lagrangian is
\begin{equation}
L=
L_0
-\tfrac12\eta(1+\frac{\eta}{z^2})J_+\bar J_-
+\tfrac12\eta J_1\bar J_0
-\tfrac18\eta^2J_0\bar J_0
-\tfrac12\eta^2\frac{x^-}{z^2}J_1\bar J_-
+\mathcal O(\eta^3)\,.
\end{equation}
The action is clearly not invariant under the isometry corresponding to constant shifts of $x^-$ and therefore the corresponding chiral current $J_+$ does not extend to the deformed theory. Instead the equations of motion lead to chiral currents
\begin{equation}
J_+^\eta=(1+\frac{\eta}{z^2})J_+-\eta\frac{x^-}{z^2}J_1+\mathcal O(\eta^2)
\,,\qquad
J_1^\eta=J_1+\mathcal O(\eta^2)\,,
\end{equation}
while the remaining equations of motion read
\begin{equation}
\partial[(1+\frac{\eta}{z^2})\bar J_-]-\eta J_1\bar J_-=\mathcal O(\eta^2)\,,\qquad
\partial[\bar J_0+\eta\frac{x^-}{z^2}\bar J_-]-\eta J_+\bar J_-=\mathcal O(\eta^2)\,.
\end{equation}
At the same time we have
\begin{equation}
\frac{dS}{d\eta}=
-\frac{T}{2}\int\,\left(J_+^\eta(1+\frac{\eta}{z^2})\bar J_--J_1\bar J_0+3\eta\frac{x^-}{z^2}J_1\bar J_-+\tfrac12\eta J_0\bar J_0+\mathcal O(\eta^2)\right)\,,
\end{equation}
which clearly cannot be written as a bilinear in deformed chiral currents.
Explicitly, we obtain the background
\footnote{Here we are writing the background that is obtained \emph{before}  doing the field redefinition.}
\be
\begin{aligned}
ds^2=&
-\frac{\eta  d\phi_1 (x^+ dx^-+x^-  dx^+)+4 dx^- dx^++\left(\eta -z^2\right) d\phi_1^2}{\eta  (\eta  x^- x^+-4)+4  z^2}
+\frac{dz^2}{z^2}\\
&+\frac{d\phi_2^2+d\phi_3^2}{4}
-\frac{2 d\phi_2 \sin \phi_3 \left(\eta  x^- dx^++\left(\eta -z^2\right) d\phi_1\right)}{\eta  (\eta  x^- x^+-4)+4 z^2},\\
B=&\frac{-4 dx^-\wedge dx^+-2 \sin \phi_3 \left( d\phi_2\wedge(\eta  x^- dx^++\left(\eta -z^2\right) d\phi_1)\right)+\eta d\phi_1\wedge(  x^+  dx^--  x^-  dx^+)}{2 \eta  (\eta  x^- x^+-4)+8 z^2},\\
e^{-2\Phi}=&1+\frac{\eta(-4+\eta x^+x^-)}{4z^2}.
\end{aligned}
\ee
{Finally, $\mathbf R_{10}=R_5+T\wedge S_++T'\wedge \bar S_-$. Since the $\alg{sl}(2,\mathbb R)$ $R$-matrix $R_5$ does not break the isometries generated by $S_+,\bar S_-$, we conclude that the additional terms in $\mathbf R_{10}$ have the only effect of adding further TsT transformations on top of the background generated by $R_5$, which is itself a TsT background.}

Interestingly, the matrices $\mathbf R_4$ (which may be decomposed in terms of $\mathbf R_1$) and $\mathbf R_9$  give rise (at leading order in the deformation parameter) to marginal deformations of the WZW model that obviously satisfy the weak CS condition, but not the strong one.

\section{Discussion}\label{sec:disc}
In this paper we have constructed YB deformations of strings on the pure NSNS $AdS_3\times S^3\times T^4$ background. Together with abelian YB deformations, which are known to reproduce TsT transformations, we also  constructed non-abelian YB deformations. While some non-abelian $R$-matrices give rise to backgrounds that cannot be obtained simply from TsT transformations, we found that  others generate again TsT backgrounds, or even no deformation at all.\footnote{Except for the introduction of a (decoupled) non-vanishing vector $K$ in the equations of generalised supergravity.} We expect this to be related to the fact that the initial $G-B$ is degenerate.

For example, the Jordanian $R$-matrix $R_1=S_0\wedge S_+$ gives back the undeformed $AdS_3$ background up to an $\eta$-dependent field redefinition (and up to a non-vanishing $K=-\eta \partial_{x^+}$). Recalling that the YB deformation is equivalent to a shift of the $B$-field plus NATD, this observation suggests that $AdS_3$ with NSNS flux has a certain property of self T-duality, when we dualise the non-abelian algebra of isometries generated by $S_0,S_+$ and we also regularise the action by performing a $B$-field gauge transformation.

Although we used our classification of $R$-matrices to deform, for concreteness, only the $AdS_3\times S^3$ part of the background, our results may also be used to obtain deformations involving the $T^4$ of $AdS_3\times S^3\times T^4$, or even deformations of the more general $AdS_3\times S^3\times S^3\times S^1$ background. Indeed, in all our expressions of the $R$-matrices the generators $T_1,\bar T_2$ may be substituted with any other two commuting generators of the compact part of the isometry algebra.\footnote{That is because the CYBE implies that the restriction of an $R$-matrix to a compact algebra must be abelian. We are therefore free to choose which abelian subalgebra we wish to consider.}
Let us also mention that the string on these $AdS_3$ backgrounds is integrable~\cite{Cagnazzo:2012se,Sundin:2012gc} and that our deformations preserve the classical integrability.

To leading order in the deformation parameter, all our YB deformations reduce to the marginal current-current deformations of the type considered by Chaudhuri and Schwartz in~\cite{Chaudhuri:1988qb}. While they are all {marginal to lowest order}, since  they satisfy what we called the ``weak CS condition'', some of them do not satisfy the ``strong CS condition'' and the celebrated ``no simple-pole condition'' {which guarantee exact marginality. Indeed we found examples which failed to be marginal beyond lowest order (and at one-loop in $1/k$), all involving the $R$-matrix $R_4$ in Table \ref{tab:non-ab-sl2}, and one example, $\mathbf R_9$ in Table \ref{tab:non-ab-sl2-su2-r4}, which remains marginal to all orders in $\eta$ at least up to one loop in $1/k$.} The relation between the space of solutions of the CYBE and that of the weak CS condition is an interesting question. The former is a quadratic equation for $R$, while the latter is a quartic equation for the coefficients $c^{ab}$ of the current-current deformation, related to the left-right block of the $R$-matrix simply as $c^{ab}=R_{\sL\sR}^{ab}$. While we expect all solutions of the CYBE to generate solutions of the weak CS condition (including the trivial ones) it seems hard to prove this statement for a generic Lie algebra. In appendix~\ref{app:sl3} we took a digression from the setup of the paper and we considered the CYBE on the $\alg{sl}_3$ algebra, finding again that it generates non-trivial solutions to the weak CS condition (that do not solve the strong one). We do not rule out the possibility of having solutions to the weak CS condition that cannot be ``completed'' to  a solution of the CYBE equation. 
In section~\ref{sec:YB} we actually discussed a more generic criterion (related to the solvability of the subalgebras involved) not requiring CYBE, to construct solutions of the weak CS condition.

In~\cite{Azeyanagi:2012zd} a marginal deformation constructed out of abelian currents (a TsT transformation) was interpreted in terms of spectral flow. It would be interesting to understand if this can be generalised to the non-abelian set-up.

We would like to stress that we worked out the YB deformations in the sigma model description. It would be very interesting to understand how to formulate the YB deformation directly at the level of the WZW action. Such a construction was performed in~\cite{Delduc:2014uaa,Delduc:2017fib,Delduc:2018xug} for $R$ a solution of the \emph{modified} CYBE.\footnote{There the special propriety $R^3=-R$ was used, so that we do not expect their results to be immediately applicable to the case of the homogeneous CYBE. Moreover, here we want $R$ to be a solution of the CYBE on $\alg f_{\sL}\oplus \alg f_{\sR}$ that also couples the left and right copy of the algebra.} The formulation of the deformation of the WZW action may be obtained from the construction in terms of NATD, in the spirit of~\cite{Hoare:2016wsk} and~\cite{Borsato:2016pas}. An alternative may be to use the language of $\mathcal E$ models~\cite{Klimcik:1995dy,Stern:1998my,Klimcik:2015gba}, see~\cite{Demulder:2018lmj} for a recent application in similar contexts.

One motivation to carry out this work came from the recent developments on the $T\bar T$ deformation~\cite{Smirnov:2016lqw,Cavaglia:2016oda} and its generalisations. The components $T,\bar T$ of the stress-energy tensor of a (quite generic) two-dimensional relativistic field theory may be used to construct a ``double-trace'' operator generating an irrelevant perturbation of the theory. The deformation is solvable in the sense that the spectrum of the deformed theory may be computed \emph{exactly} in the deformation parameter  as a function of the spectrum of the original undeformed theory. In~\cite{Giveon:2017nie} a ``single-trace'' version of the $T\bar T$ deformation of the symmetric product orbifold CFT was considered. It was argued that the irrelevant deformation of the ``spacetime'' CFT governed by\footnote{We refer to~\cite{Giveon:2017nie} for the connection to Little String Theories. The above operator should be compared to the original double-trace version studied in~\cite{Smirnov:2016lqw,Cavaglia:2016oda} and given by $T(x)\bar T(\bar x)$, where $T(x)=\sum_{i=1}^NT_i(x)$ and $\bar T(\bar x)=\sum_{i=1}^N\bar T_i(\bar x)$.} $\mathcal O(x)\propto \sum_{i=1}^NT_i(x)\bar T_i(\bar x)$, where $i$ labels each copy in the symmetric product,  corresponds to a marginal deformation of the dual WZW model that infinitesimally is just the current-current deformation  $J_+(\sigma)\bar J_-(\bar \sigma)$, where $J_+,\bar J_-$ are the left and right $SL(2,\mathbb R)$ currents generating shifts of the boundary coordinates $x^+,x^-$. 
Another deformation, similar in spirit to the above one, was studied in~\cite{Chakraborty:2018vja} and~\cite{Apolo:2018qpq} after replacing the $\bar T$ with an antichiral $U(1)$ current of the compact factor.\footnote{This   deformation is in fact the single-trace version of the one first constructed in~\cite{Guica:2017lia}.}
It was argued in~\cite{Chakraborty:2018vja,Apolo:2018qpq} that the deformation of the dual WZW model is governed again by a marginal deformation bilinear in the currents (where now the antichiral current belongs to the compact part of the algebra). Such marginal deformations of the WZW model may be completed to finite values of the deformation parameter in terms of TsT (or equivalently certain $O(d,d)$) transformations. 
Since TsT deformations are a subclass of YB ones, it would be interesting to understand if it is possible to provide a holographic interpretation also for the YB deformations of $AdS_3\times S^3\times T^4$ considered here. (The connection to YB models was also pointed out the recent paper~\cite{Araujo:2018rho}.) We expect our marginal deformations of the WZW model to correspond to deformations of the dual CFT$_2$ which generalise the (single trace version of the) $T\bar T$ construction. 
It would be very interesting to understand for example the case of the non-abelian $R$-matrix $\mathbf R_9$, which gives rise to the marginal deformation $aJ_0\bar J_2+bJ_1\bar J_0+cJ_1\bar J_2+J_+\bar J_-$. The non-abelianity of the generators involved forbids the usual iteration of the infinitesimal deformation in order to obtain the exact one. The YB deformation, despite the non-abelianity, provides the realisation of the finite deformation on the worldsheet of the string.

\section*{Acknowledgements}
We thank  R. Conti, B. Hoare and D. Thompson for useful discussions. LW thanks the participants of the workshop ``A fresh look at $AdS_3/CFT_2$'' in Villa Garbald, Castasegna, for stimulating discussions. The work of RB is supported by the Maria de Maeztu Unit of Excellence MDM-2016-0692, by FPA2017-84436-P, by Xunta de Galicia (ED431C 2017/07), and by FEDER. His work was supported by the ERC advanced grant No 341222 while at Nordita.

\vspace{2cm}

\appendix

\section{Details on the field redefinition}\label{app:field-red}
The matrix $M$ in~\eqref{eq:Lprime} is a direct sum of the $AdS$ and sphere part, $M=M_{\rm a}\oplus M_{\rm s}$. They take the form (we restrict to $A,B=\{0,+,\bar-,\bar0\}$ and $A,B=\{1,\bar2\}$ respectively, which is all we need since $M$ always comes multiplied with $R$ on both sides)
\begin{equation}
M_{\rm a}=
\left(
\begin{array}{cccc}
\tfrac14 & 0 & 0 & -\tfrac14\\
0 & 0 & 0 & 0\\
-\frac{x^+}{z^2} & -\frac{1}{z^2} & 0 & 0\\
-\tfrac14+\frac{x^+x^-}{z^2} & \frac{x^-}{z^2} & 0 & \tfrac14\\
\end{array}
\right)\,,
\qquad
M_{\rm s}=
\left(
\begin{array}{cc}
\tfrac14 & 0\\
-\tfrac12\sin{\phi_3} & \tfrac14
\end{array}
\right)\,.
\label{eq:M}
\end{equation}

We will make an ansatz for the field redefinition based on the isometry transformations used to construct the model, but with the transformation parameters depending linearly on $y=(\ln z,\phi_1,\phi_2)$ (since we want to cancel terms involving $\partial y$ and $\bar\partial y$). We therefore consider
\begin{equation}
x^\pm\rightarrow x'^\pm=e^{\eta b^i_\pm y_i}[x^\pm+\eta a^i_\pm y_i]\,,\qquad
z\rightarrow z'=e^{\eta b^iy_i}z\,,\qquad
\phi_{1,2}\rightarrow\phi'_{1,2}=\phi_{1,2}+\eta a_{1,2}^iy_i\,,
\end{equation}
where $i=1,2,3$ and $b^i=(b^i_++b^i_-)/2$. For the right-moving Noether currents $\mathcal J_{A+}$ we get
\begin{align}
\mathcal J'_{0+}=&J_0
+\eta a^i_+y_iJ_+
-\tfrac12\partial\ln z
+\partial y_i[\eta b_+^iM'_{00}-\eta b^i_-M'_{\bar00}+\eta a^i_-e^{\eta b^i_-y_i}M'_{\bar-0}]
\,,\\
\mathcal J'_{++}=&e^{-\eta b^i_+y_i}J_+
+\partial y_i[\eta a^i_-e^{\eta b_-^iy_i}M'_{\bar-+}-\eta b^i_-M'_{\bar0+}]
\,,\\
\mathcal J'_{\bar0+}=&-\tfrac12\partial\ln z-\tfrac12\eta b^i\partial y_i
\,,\\
\mathcal J'_{1+}=&J_1+\tfrac14\partial\phi_1
+\partial y_i[\eta a_2^iM'_{\bar21}-\eta a_1^iM_{11}]
\,,\\
\mathcal J'_{\bar2+}=&\tfrac14\partial\phi_2+\tfrac14\eta a_2^i\partial y_i\,.
\end{align}
These expressions take a more natural form if we further assume that
\begin{equation}
a_1^i=-Y^i_AR^{A1}\,,\quad
a_2^i=Y^i_AR^{A\bar2}\,,\quad
a_+^i=Y^i_AR^{A+}\,,\quad
a_-^i=Y^i_AR^{A\bar-}\,,\quad
b_+^i=Y^i_AR^{A0}\,,\quad
b_-^i=-Y^i_AR^{A\bar0}\,,
\end{equation}
for some constants $Y_A^i$ to be determined. Then we have
\begin{equation}
\mathcal J'_{A+}=e^{\delta_A}J_A+\partial y_iY^i_B[\eta RM']^B{}_A+\Delta_A
\end{equation}
with $\delta_+=-\eta Y^i_AR^{A0}y_i$ and 
\begin{align}
\Delta_0=&-\tfrac12\partial\ln z+\eta y_iY^i_AR^{A+}J_++[e^{\eta b^i_-y_i}-1]\partial y_iY^i_A\eta R^{A\bar-}M'_{\bar-0}
\,,\\
\Delta_+=&[e^{\eta b_-^iy_i}-1]\partial y_iY^i_A\eta R^{A\bar-}M'_{\bar-+}
\,,\\
\Delta_{\bar0}=&-\tfrac12\partial\ln z\,,\qquad
\Delta_1=\tfrac14\partial\phi_1\,,\qquad
\Delta_{\bar2}=\tfrac14\partial\phi_2\,,
\end{align}
the other components vanishing. Using these expressions the transformed Lagrangian becomes
\begin{equation}
L'=
L'_0
-\tfrac12\eta e^{\delta_a}J_a[(1+\eta RM')^{-1}R]^{aB}\mathcal J'_{B-}
+\tfrac12\eta(\partial y_iY^i_A-\Delta_A)[(1+\eta RM')^{-1}R]^{AB}\mathcal J'_{B-}
-\tfrac12\eta\partial y_iY^i_AR^{AB}\mathcal J'_{B-}\,.
\end{equation}
Picking $Y^i_A$ such that $\partial y_iY^i_A-\Delta_A$ vanishes to lowest order in $\eta$ we get that its non-zero components should be
\begin{equation}
Y^1_0=Y^1_{\bar0}=-\frac12\,,\qquad
Y^2_1=Y^3_{\bar2}=\frac14
\end{equation}
and the Lagrangian becomes
\begin{equation}
L'=
L'_0
-\tfrac12\eta\partial y_iY^i_AR^{AB}\mathcal J'_{B-}
-\tfrac12\eta\hat J_a[(1+\eta RM')^{-1}R]^{aB}\mathcal J'_{B-}\,.
\end{equation}
where we have defined
\begin{equation}
\hat J_a=e^{\delta_a}J_a+\eta\delta_a^0y_iY^i_AR^{A+}J_++\eta[e^{-\bar\delta_-}-1]\partial y_iY^i_AR^{A\bar-}M'_{\bar-a}\,,
\label{eq:Jhat}
\end{equation}
with $\bar\delta_-=\eta Y^i_AR^{A\bar0}y_i$. A short calculation shows that the transformed undeformed Lagrangian is (up to total derivatives)
\begin{align}
L'_0
=&
L_0
+\tfrac12\eta Y^i_AR^{Aa}\left(J_a+\eta\delta_a^0y_jY^j_BR^{B+}J_+\right)\bar\partial y_i
+\tfrac12\eta\partial y_iY^i_AR^{A\bar a}\bar J'_{\bar a}
\nonumber\\
&{}
+\tfrac12\eta[e^{-\bar\delta_-}-1]Y^i_AR^{A\bar-}\partial y_i\bar J_-'
-\tfrac12\eta\partial y_iY^i_AR^{AB}Y^j_B\bar\partial y'_j
\end{align}
and using this together with the fact that $\mathcal J_{B-}=\bar J_B-Y^i_B\bar\partial y_i$ we get
\begin{align}
L'=&
L_0
+\tfrac12\eta Y^i_AR^{Aa}(J_a+\eta\delta_a^0y_jY^j_BR^{B+}J_+)\bar\partial y_i
-\tfrac12\eta\hat J_a[(1+\eta RM')^{-1}R]^{aB}\mathcal J'_{B-}
\nonumber\\
&{}
+\tfrac12\eta[e^{\eta b^i_-y_i}-1]\partial y_iY^i_AR^{A\bar-}\mathcal J'_{\bar--}\,.
\end{align}
Finally we use the fact that
\begin{equation}
\mathcal J'_{B-}=e^{\bar\delta_B}\bar J_B-[1+\eta M'R]_B{}^AY^i_A\bar\partial y_i+\bar\Delta_B
\end{equation}
with $\bar\delta_{\bar-}=\bar\delta_-$ was defined above and
\begin{align}
\bar\Delta_{\bar-}=&-\eta[e^{-\delta_+}-1]M'_{\bar-+}R^{+A}Y^i_A\bar\partial y_i
\,,\\
\bar\Delta_{\bar0}=&-\eta Y^i_AR^{A\bar-}y_i\bar J_--\eta[e^{-\delta_+}-1]Y^i_AM'_{\bar0+}R^{+A}\bar\partial y_i\,,
%
\end{align}
and the remaining components vanishing. Finally the Lagrangian becomes
\begin{align}
L'=&
L_0
-\tfrac12\eta\hat J_a[(1+\eta RM')^{-1}R]^{a\bar a}\hat{\bar J}_{\bar a}
+\tfrac12\eta[e^{\delta_+}-1]J_+R^{+A}Y^i_A\bar\partial y_i
-\tfrac12\eta[e^{\bar\delta_-}-1]\partial y_iY^i_AR^{A\bar-}\bar J_-
\nonumber\\
&{}
-\tfrac12\eta^2[e^{-\bar\delta_-}-1]\partial y_iY^i_AR^{A\bar-}M'_{\bar-+}[e^{-\delta_+}-1]Y^j_BR^{+B}\bar\partial y_j\,,
\label{eq:Ldef}
\end{align}
with (recall that $\delta_+=-\eta Y^i_AR^{A0}y_i$ and $\bar\delta_{\bar-}=\eta Y^i_AR^{A\bar0}y_i$)
\begin{equation}
\hat{\bar J}_{\bar a}=e^{\bar\delta_{\bar a}}\bar J_{\bar a}-\eta\delta_{\bar a}^{\bar0}Y^i_AR^{A\bar-}y_i\bar J_--\eta M'_{\bar a+}[e^{-\delta_+}-1]R^{+A}Y^i_A\bar\partial y_i\,.
\label{eq:Jbarhat}
\end{equation}
The terms involving $e^{\delta_+}-1$ and $e^{\bar\delta_-}-1$ are not expressed in terms of the chiral currents but they only appear at order $\eta^2$, so they do not interfere with the comparison to infinitesimal current-current deformations. In fact they vanish for most of the deformations of $AdS_3\times S^3$, e.g. for $AdS_3$ deformations we have $R^{0\bar0}R^{A+}=R^{0\bar0}R^{A\bar-}=0$ and using this we find
\begin{equation}
L=
L_0
-\tfrac12\eta\hat J_a[(1+\eta RM')^{-1}R]^{a\bar a}\hat{\bar J}_{\bar a}\,,
\label{eq-app:LdefAdS3}
\end{equation}
with
\begin{equation}
\hat J_a=J_a-\tfrac12\eta\delta_a^0(R^{0+}+R^{\bar0+})\ln zJ_+\,,\qquad
\hat{\bar J}_{\bar a}=\bar J_{\bar a}+\tfrac12\eta\delta_{\bar a}^{\bar0}(R^{0\bar-}+R^{\bar0\bar-})\ln z\bar J_-\,.
\label{eq:JhatAdS3}
\end{equation}

\section{On-shell equivalence}~\label{app:onshell}
Here we will demonstrate the on-shell equivalence of the YB deformed sigma models to the original ones by deriving the explicit (non-local) field redefinition that relates them. We will do that by following the NATD transformation and following field redefinition that are used to get the action of the YB model as in~\cite{Borsato:2018idb}.
In the notation of~\cite{Borsato:2018idb}, let us start from the action 
\begin{align}
S'=&\frac{T}{2}\int_\Sigma\,\Big(
A^I\wedge(G_{IJ}*-B_{IJ})A^J
+2dz^m\wedge(G_{mI}*-B_{mI})A^I
\nonumber\\
&\qquad{}
+dz^m\wedge(G_{mn}*-B_{mn})dz^n
\Big)\,,
\label{eq:S-first}
\end{align}
where $A=g^{-1}d g$ and we have set fermions to zero for simplicity. When going to the NATD model one relates the original degrees of freedom to the new ones encoded in the Lagrange multiplier $\nu$ as
\begin{equation}\label{eq:integroutA}
(1\pm*)A^I=-(1\pm*)\left(d\nu_J+dz^m[\mp G-B]_{mJ}\right)N_\mp^{JI}\,,\qquad
N_\pm^{IJ}=\left(\pm G_{IJ}-B_{IJ}-\nu_Kf^K_{IJ}\right)^{-1}.
\end{equation}
The YB model appears after the redefinition\footnote{It is assumed that the initial $B$-field is actually shifted as $B_{IJ}\to B_{IJ}-\eta^{-1}R^{-1}_{IJ}$.}
\be
\nu_I = \eta^{-1}\tr\left(T_I \frac{1-\Ad_{\tilde g}^{-1}}{\log \Ad_{\tilde g}}R^{-1} \log {\tilde g}\right)\,,
\ee
which implies
\be
d\nu_I=\eta^{-1}\left(R_{\tilde g}^{-1}(\tilde g^{-1}d\tilde g)\right)_I\,,\qquad
N=\eta R_{\tilde g}\left(1+\eta(G-B)R_{\tilde g}\right)^{-1}=\eta \left(1+\eta R_{\tilde g}(G-B)\right)^{-1}R_{\tilde g}\,,
\ee
where $N=N_+=-N_-^T$.
Combining all redefinitions we find that the original $A=g^{-1}d g$ (which will depend on some coordinates $x^i$) is related to the new degrees of freedom of the YB model (i.e. the coordinates $\tilde x^i$ parameterising $\tilde g$, together with the coordinates $z^m$ that remain spectators) as
\be
(1\pm *)A^I=(1\pm *)[(1\pm \eta R_{\tilde g}(G\mp B))^{-1}]^{ I}_{J}
\left( (\tilde g^{-1}d\tilde g)^J\mp \eta R_{\tilde g}^{JK}(G\mp B)_{Km}dz^m \right).
\ee 
Using formula (4.21) of~\cite{Borsato:2018idb} we find that it can be written as
\be
(1\pm *)A=(1\pm *)(1\pm \eta R_{\tilde g}(G\mp B))^{-1}\tilde V,
\qquad
\tilde V^M\equiv  \delta^M_I(\tilde g^{-1}d\tilde g)^I+\delta^M_mdz^m.
\ee
This can \emph{almost} be written as a relation involving only the derivatives of the redefined coordinates
\be
(1\pm *)W=(1\pm *)(1\pm \eta \Theta(G\mp B))^{-1}d\tilde X,
\qquad
d\tilde X^M\equiv  \delta^M_id\tilde x^i+\delta^M_mdz^m,
\ee
where
\be
W^M= \delta^M_i\ell^i_I\tilde \ell^I_jdx^j+\delta^M_mdz^m,
\qquad
g^{-1}dg=dx^i\ell_i^IT_I,\quad
\tilde g^{-1}d\tilde g=d\tilde x^i\tilde \ell_i^IT_I
\ee
and where all indices that have been omitted above are curved indices $M=\{i,m\}$.
In general $\ell^i_I\tilde \ell^I_j\neq \delta^i_j$ and $G,B,\Theta$ may depend on $\tilde x$.
Thanks to the above field redefinition we can argue that solutions to the classical equations of the YB model can be mapped to solutions of the undeformed model, and vice versa.

\section{All non-abelian $R$-matrices of $\mathfrak{sl}(2,\mathbbm R)^2\oplus\mathfrak{su}(2)^2$}\label{app:R-matrices}
We will classify non-abelian $R$-matrices solving the classical Yang-Baxter equation relevant to deformations of $AdS_3\times S^3$. The $R$-matrix is an anti-symmetric matrix with indices in the isometry algebra, in our case
\begin{equation}
\mathfrak{sl}(2,\mathbbm R)_{\sL}\oplus\mathfrak{sl}(2,\mathbbm R)_{\sR}\oplus\mathfrak{su}(2)_{\sL}\oplus\mathfrak{su}(2)_{\sR}\,.
\end{equation}
The classical Yang-Baxter equation (CYBE)
\begin{equation}
R^{[A|B|}R^{C|D|}f^{E]}_{BD}=0\,,
\end{equation}
implies that $R^{AB}$ is non-degenerate on a subalgebra and zero elsewhere. Calling the inverse $\omega_{AB}$ the CYBE is equivalent to $\omega_{A[B}f^A_{CD]}=0$, i.e. $\omega$ is a Lie algebra 2-cocycle on the (dual of the) subalgebra where $R$ defined. Since it is also invertible this subalgebra is a quasi-Frobenius (sometimes also called symplectic) subalgebra.\footnote{If $\omega$ is exact, i.e. $\omega_{AB}=f_{AB}^CX_C$ for some $X_C$, the algebra is Frobenius.} Therefore $R$-matrices solving the CYBE on some Lie algebra are in one-to-one correspondence with quasi-Frobenius subalgebras of this Lie algebra~\cite{STOLIN1999285}.

Semi-simple Lie algebras cannot be quasi-Frobenius and therefore such algebras of dimension 4 (or 2) must be solvable \cite{zbMATH03411826}. In our case things are particularly simple since we have a sum of four 3 dimensional Lie algebras. When can therefore restrict our attention to subalgebras of the maximal solvable subalgebra of the isometry algebra which we take to be
\begin{equation}
\mathfrak s=\mathrm{span}\{S_0,\,S_+,\,\bar S_0,\,\bar S_-,\,T_1,\,\bar T_2\}\,.
\end{equation}

The only non-abelian solvable Lie algebra of dimension 2 is $\mathfrak r_2$ with Lie bracket $[e_1,e_2]=e_2$. The possible $R$-matrices of rank 2 are given by embeddings of $\mathfrak r_2$ into $\mathfrak s$ and up to $SL^2(2)\times SU^2(2)$ automorphisms they are
\begin{equation}
\begin{array}{l}
\mathbf R_1=(S_0-\bar S_0+T)\wedge(S_+\pm\bar S_-)\\
\mathbf R_2=(S_0-a\bar S_0+T)\wedge S_+\\
\mathbf R_3=(S_0\mp\bar S_-+T)\wedge S_+
\end{array}
\end{equation}
where $T$ is any linear combination of $T_1$ and $\bar T_2$.

For the rank 4 case we need to find 4-dimensional solvable subalgebras of $\mathfrak s$, which are furthermore quasi-Frobenius (symplectic). The complete list of 4-dimensional quasi-Frobenius algebras can be found in \cite{Ovando2006}. Taking into account the fact that $[\mathfrak s,\mathfrak s]=\mathrm{span}\{S_+,\,\bar S_-\}$ is 2-dimensional we can rule out any algebra with more than two independent linear combinations of generators arising from commutators. It is also trivial to see that the Heisenberg algebra with non-trivial Lie bracket $[e_1,e_2]=e_3$ is not a subalgebra, $\mathfrak h_3\not\subset\mathfrak s$. Using these two facts the list of 4-dimensional solvable subalgebras is reduced to (note that $\mathfrak r_{4,-1,0}=\mathbbm{R}\oplus\mathfrak r_{3,-1}$)
\begin{equation}
\mathbbm{R}\oplus\mathfrak r_3\,,\quad
\mathbbm{R}\oplus\mathfrak r_{3,\lambda}\,,\quad
\mathbbm{R}\oplus\mathfrak r'_{3,\gamma}\,,\quad
\mathfrak r_2\oplus\mathfrak r_2\,,\quad
\mathfrak r'_2\,,\quad
\mathfrak d_{4,1}\,.
\end{equation}
The same paper lists the symplectic ones which are
\begin{equation}
\begin{array}{l|l}
\mathbbm{R}\oplus\mathfrak r_{3,0} & [e_1,e_2]=e_2\\
\mathbbm{R}\oplus\mathfrak r_{3,-1} & [e_1,e_2]=e_2\,,\quad [e_1,e_3]=-e_3\\
\mathbbm{R}\oplus\mathfrak r'_{3,0} & [e_1,e_2]=-e_3\,,\quad [e_1,e_3]=e_2\\
\mathfrak r_2\oplus\mathfrak r_2 & [e_1,e_2]=e_2\,,\quad [e_3,e_4]=e_4\\
\mathfrak r'_2 & [e_1,e_3]=e_3\,,\quad [e_1,e_4]=e_4\,,\quad [e_2,e_3]=e_4\,,\quad [e_2,e_4]=-e_3\\
\mathfrak d_{4,1} & [e_1,e_2]=e_3\,,\quad [e_4,e_3]=e_3\,,\quad [e_4,e_1]=e_1
\end{array}
\end{equation}
It is not hard to show, using the fact that the elements arising from commutators are of the form $aS_++b\bar S_-$ for some $a,b$, that $\mathfrak r'_{3,0}$ is not a subalgebra of $\mathfrak s$ and therefore neither is $\mathfrak r'_2$. It is also not hard to show that neither is $\mathfrak d_{4,1}$. For the remaining ones we find the embeddings (again up to automorphism)
\begin{equation}
\begin{array}{l|l}
\mathbbm{R}\oplus\mathfrak r_{3,0} & e_1=S_0-b\bar S_0+e_1'\,,\quad e_2=S_++a\bar S_-\qquad (a=0\quad\mbox{or}\quad a=b=1)\\
\mathbbm{R}\oplus\mathfrak r_{3,-1} & e_1=S_0+\bar S_0+e_1'\,,\quad e_2=S_+\,,\quad e_3=\bar S_-\\
\mathfrak r_2\oplus\mathfrak r_2 & e_1=S_0+e_1'\,,\quad e_2=S_+\,,\quad e_3=-\bar S_0+e_3'\,,\quad e_4=\bar S_-\\
\end{array}
\end{equation}
Primed generators denote any linear combination that commutes with the remaining generator. Only the second algebra is unimodular (i.e. the trace of its structure constants vanish) and is contained in the classification of unimodular $R$-matrices in \cite{Borsato:2016ose}.

The rank 4 $R$-matrices can then be read off from the classification in \cite{Ovando2006}. Up to inner automorphisms and exchanging the left and right copy of the algebra they are
\begin{equation}
\begin{array}{l}
\mathbf R_4=(S_0-\bar S_0+T)\wedge(S_+\pm\bar S_-)+aT_1\wedge\bar T_2\\
\mathbf R_5=(S_0-a\bar S_0+T)\wedge S_++bT_1\wedge\bar T_2\\
\mathbf R_6=(S_0+T)\wedge S_++T'\wedge(\bar S_0+T'')\\
\mathbf R_7=(S_0+T)\wedge S_++T'\wedge(\bar S_-+T'')\\
\mathbf R_8=(S_0\mp\bar S_-+T)\wedge S_++aT_1\wedge\bar T_2\\
\mathbf R_9=(S_0+\bar S_0+T)\wedge T'+S_+\wedge\bar S_-\\
\mathbf R_{10}={(S_0+T)\wedge S_+\pm(\bar S_0+T')\wedge \bar S_-+\lambda S_+\wedge \bar S_-}\\
\end{array}
\end{equation}
where $T,T',T''$ are linear combinations of $T_1$ and $\bar T_2$. The rank 2 $R$-matrices are contained as special cases of $\mathbf R_4,\mathbf R_5,\mathbf R_8$ when the last term vanishes.

A rank 6 $R$-matrix is only possible if $\mathfrak s$ is itself quasi-Frobenius. It is in this case but the resulting $R$-matrix is just $\mathbf R_{10}+aT_1\wedge\bar T_2$ and therefore leads just to a TsT transformation of the $\mathbf R_{10}$ deformation. Therefore we will not consider it further here.

\section{$R$-matrix on parabolic subalgebra of $\alg{sl}_3$}\label{app:sl3}
Let $\mathfrak f$ denote the 6-dimensional parabolic Lie subalgebra of $\mathfrak{sl}_3$ \cite{Stolin:1991} with basis (see example 3.6 of \cite{Burde:2016})
\begin{equation}
(e_1,\ldots,e_6)=(E_{12},E_{13},E_{21},E_{23},E_{11}-E_{22},E_{22}-E_{33})\,.
\end{equation}
The Lie brackets are given by
\begin{align}
&{}[e_1,e_3]=e_5\,,\quad
[e_1,e_4]=e_2\,,\quad
[e_1,e_5]=-2e_1\,,\quad
[e_1,e_6]=e_1\,,\quad
[e_2,e_3]=-e_4\,,\quad
\nonumber\\
&{}[e_2,e_5]=-e_2\,,\quad
[e_2,e_6]=-e_2\,,\quad
[e_3,e_5]=2e_3\,,\quad
[e_3,e_6]=-e_3\,,\quad
[e_4,e_5]=e_4\,,\quad
[e_4,e_6]=-2e_4\,.
\end{align}
This algebra is not unimodular since $f_{i6}{}^i=-3\neq0$ and also not solvable since $\mathrm{tr}([e_1,e_3]e_5)\neq0$ but it is quasi-Frobenius with 2-cocycle
\begin{align}
\omega=&
a_{13}e^1\wedge e^3
+a_{14}e^1\wedge e^4
-2a_{16}e^1\wedge e^5
+a_{16}e^1\wedge e^6
+a_{23}e^2\wedge e^3
-a_{14}e^2\wedge e^5
-a_{14}e^2\wedge e^6
\nonumber\\
&{}
-2a_{36}e^3\wedge e^5
+a_{36}e^3\wedge e^6
-a_{23}e^4\wedge e^5
+2a_{23}e^4\wedge e^6.
\end{align}
One may take e.g.
\begin{equation}
\omega=
e^1\wedge(2e^5-e^6)
+e^2\wedge e^3
+e^4\wedge(2e^6-e^5)
\end{equation}
the inverse being
\begin{equation}
R=
-\frac13e_1\wedge(2e_5+e_6)
-e_2\wedge e_3
-\frac13e_4\wedge(2e_6+e_5)\,.
\end{equation}
For the $SL(3)$ WZW model we can get a deformation by embedding this algebra in the diagonal $SL(3)\subset SL(3)\times SL(3)$. 
After identifying the coefficients of CS with the off-diagonal components of the $R$-matrix and using the definitions in~\eqref{eq:def-C-Cbar} we find
\begin{align}
&{}
C^{12\bar 3}=-1\,,\quad
C^{13\bar 2}=-1\,,\quad
C^{15\bar 1}=-\frac23\,,\quad
C^{16\bar 1}=-\frac13\,,\quad
C^{23\bar 4}=1\,,\quad
C^{25\bar 5}=\frac23\,,\quad
\nonumber\\
&{}
C^{26\bar 5}=\frac13\,,\quad
C^{34\bar 2}=1\,,\quad
C^{45\bar 4}=-\frac13\,,\quad
C^{46\bar 4}=-\frac23\,,\quad
C^{56\bar 2}=\frac13\,.
\end{align}
Once again we are using a bar for indices in the right copy of the algebra.
Using the fact that the non-zero components of the Killing metric are
\begin{equation}
K_{13}=K_{31}=1\,,\quad
K_{55}=K_{66}=2\,,\quad
K_{56}=K_{65}=-1\,,
\end{equation}
one finds that $C^{ab\bar c}K_{\bar c\bar d}C^{ef\bar d}\neq0$ but the weak CS condition is satisfied since $C^2=0$.

\section{More general ``trivial'' solutions of generalized supergravity}\label{app:trivial}
In \cite{Wulff:2018aku} it was shown that the generalized supergravity equations can have ``trivial'' solutions, i.e. ones that are also solutions of standard supergravity even though the Killing vector $K$ is non-zero. Writing\footnote{We recall that this was denoted just by $X$ in \cite{Wulff:2018aku}, but we prefer to avoid confusion with the vector $X={\rm X}+K$ of~\cite{Wulff:2016tju}.} ${\rm X}=d\phi+i_KB$ it was shown that for $\phi$ to have a gauge invariant meaning $K$ must be null. While this is a natural condition it is not strictly necessary. It is possible to have a situation where $K$ is not null that still leads to a standard supergravity solution, as we will show here. However in that case one must pick the correct gauge for the $B$-field to read off the dilaton from the expression ${\rm X}=d\phi+i_KB$, as in a different gauge one may find a different $\phi$ which will not be the correct dilaton of a standard supergravity solution. {To avoid having to deal with gauge-transformations of the B-field we will write ${\rm X}=d\phi+\tilde K$ where $\phi$ should be identified with the dilaton. Below we derive the conditions on $\tilde K$ for a trivial solution of the supergravity equations. Note that if we pick a gauge so that $B$ is invariant under the isometry generated by $K$ we have $\tilde K=i_KB+d\phi'$ for some $\phi'$.}

{
We will ignore the RR fields in our discussion. Looking at the generalized supergravity equations in \cite{Wulff:2018aku} it follows from the generalized Einstein equation that for {$(G,H,\phi)$ to solve the standard supergravity $\tilde K$ must be a Killing vector (of the metric, other fields do not a priori have to be invariant). The remaining equations lead to the following conditions to have a trivial solution
\begin{align}
&d\tilde K+i_KH=0\,,\qquad
\mathcal L_K\phi+K\cdot\tilde K=0\,,\nonumber\\
&dK+i_{\tilde K}H=0\,,\qquad
\mathcal L_{\tilde K}\phi+K^2+\tilde K^2=0\,.
\label{eq:trivial}
\end{align}
}
If $\tilde K$ is proportional to $K$ we get precisely the solutions considered in \cite{Wulff:2018aku}, {in particular $K$ is null.} But it can also happen that $\tilde K$ and $K$ are linearly independent, as the example in (\ref{eq:lambda0}) shows. Such solutions are clearly much less generic than the solutions considered in \cite{Wulff:2018aku} since they require at least two Killing vectors. They are also harder to identify since they require first extracting the correct dilaton and $\tilde K$ and then verifying the equations above.

In \cite{Wulff:2018aku} it was argued that the analysis based on the generalized supergravity equations agrees with what one gets by looking at the non-local terms in the sigma model action induced by non-abelian T-duality on a non-unimodular group \cite{Alvarez:1994np,Elitzur:1994ri}. The analysis was done for standard YB sigma models where the more general possibility of an independent Killing vector $\tilde K$ does not arise. However, the general form of the non-local terms proposed there, namely
\begin{equation}
L_\sigma=\alpha'd\sigma\wedge K-\alpha'd\sigma\wedge*{\rm X}+\mathcal O(\alpha'^2)\,,
\end{equation}
is consistent with the general analysis above since, up to total derivatives, this is equal to
\begin{equation}
{\alpha'}\sigma(dK+i_{\tilde K}H)
-{\alpha'}\sigma(d*\tilde K+i_{\tilde K}H)
{+\alpha'\phi R^{(2)}}
\end{equation}
and the first term vanishes by (\ref{eq:trivial}) and the second is proportional to the equations of motion projected along the Killing vector $\tilde K$. The order $\alpha'^2$ terms were also considered in \cite{Wulff:2018aku} and, given the present analysis, these will be modified so that they now involve a combination of the second and last equation in (\ref{eq:trivial}) rather than just $K^2$.
}

\bibliographystyle{nb}
\bibliography{biblio}{}

\begin{thebibliography}{10}
\ifx\href\asklfhas\newcommand{\href}[2]{#2}\fi
\ifx\arxivref\asklfhas\newcommand{\arxivref}[2]{\href{http://arxiv.org/abs/#1}{#2}}\fi
\ifx\doiref\asklfhas\newcommand{\doiref}[2]{\href{http://dx.doi.org/#1}{#2}}\fi
\raggedright
\small
\parskip 0pt

\bibitem{Maldacena:1997re}
J.~M.~Maldacena,
\textit{``{The Large N limit of superconformal field theories and
  supergravity}''},
\textsf{\doiref{10.1023/A:1026654312961}{Int.J.Theor.Phys.~38,~1113~(1999)}},
\texttt{\arxivref{hep-th/9711200}{hep-th/9711200}}.

\bibitem{Giveon:1998ns}
A.~Giveon, D.~Kutasov and N.~Seiberg,
\textit{``{Comments on string theory on AdS$_3$}''},
\textsf{Adv.Theor.Math.Phys.~2,~733~(1998)},
\texttt{\arxivref{hep-th/9806194}{hep-th/9806194}}.

\bibitem{Kutasov:1999xu}
D.~Kutasov and N.~Seiberg,
\textit{``{More comments on string theory on AdS(3)}''},
\textsf{\doiref{10.1088/1126-6708/1999/04/008}{JHEP~9904,~008~(1999)}},
\texttt{\arxivref{hep-th/9903219}{hep-th/9903219}}.

\bibitem{Maldacena:2000hw}
J.~M.~Maldacena and H.~Ooguri,
\textit{``{Strings in AdS$_3$ and SL(2,R) WZW model 1.: The Spectrum}''},
\textsf{\doiref{10.1063/1.1377273}{J.Math.Phys.~42,~2929~(2001)}},
\texttt{\arxivref{hep-th/0001053}{hep-th/0001053}}.

\bibitem{Maldacena:2000kv}
J.~M.~Maldacena, H.~Ooguri and J.~Son,
\textit{``{Strings in AdS$_3$ and the SL(2,R) WZW model. Part 2. Euclidean
  black hole}''},
\textsf{\doiref{10.1063/1.1377039}{J.Math.Phys.~42,~2961~(2001)}},
\texttt{\arxivref{hep-th/0005183}{hep-th/0005183}}.

\bibitem{Maldacena:2001km}
J.~M.~Maldacena and H.~Ooguri,
\textit{``{Strings in AdS(3) and the SL(2,R) WZW model. Part 3. Correlation
  functions}''},
\textsf{\doiref{10.1103/PhysRevD.65.106006}{Phys.~Rev.~D65,~106006~(2002)}},
\texttt{\arxivref{hep-th/0111180}{hep-th/0111180}}.

\bibitem{Eberhardt:2018ouy}
L.~Eberhardt, M.~R.~Gaberdiel and R.~Gopakumar,
\textit{``{The Worldsheet Dual of the Symmetric Product CFT}''},
\texttt{\arxivref{1812.01007}{arxiv:1812.01007}}.

\bibitem{Chaudhuri:1988qb}
S.~Chaudhuri and J.~A.~Schwartz,
\textit{``{A Criterion for Integrably Marginal Operators}''},
\textsf{\doiref{10.1016/0370-2693(89)90393-6}{Phys.~Lett.~B219,~291~(1989)}}.

\bibitem{Chaudhuri:1992yca}
S.~Chaudhuri and J.~D.~Lykken,
\textit{``{String theory, black holes, and SL(2,R) current algebra}''},
\textsf{\doiref{10.1016/0550-3213(93)90267-S}{Nucl.~Phys.~B396,~270~(1993)}},
\texttt{\arxivref{hep-th/9206107}{hep-th/9206107}}.

\bibitem{Hassan:1992gi}
S.~F.~Hassan and A.~Sen,
\textit{``{Marginal deformations of WZNW and coset models from O(d,d)
  transformation}''},
\textsf{\doiref{10.1016/0550-3213(93)90429-S}{Nucl.~Phys.~B405,~143~(1993)}},
\texttt{\arxivref{hep-th/9210121}{hep-th/9210121}}.

\bibitem{Kiritsis:1993ju}
E.~Kiritsis,
\textit{``{Exact duality symmetries in CFT and string theory}''},
\textsf{\doiref{10.1016/0550-3213(93)90428-R}{Nucl.~Phys.~B405,~109~(1993)}},
\texttt{\arxivref{hep-th/9302033}{hep-th/9302033}}.

\bibitem{Tseytlin:1993hm}
A.~A.~Tseytlin,
\textit{``{On A 'Universal' class of WZW type conformal models}''},
\textsf{\doiref{10.1016/0550-3213(94)90243-7}{Nucl.~Phys.~B418,~173~(1994)}},
\texttt{\arxivref{hep-th/9311062}{hep-th/9311062}}.

\bibitem{Forste:2003km}
S.~Forste and D.~Roggenkamp,
\textit{``{Current current deformations of conformal field theories, and WZW
  models}''},
\textsf{\doiref{10.1088/1126-6708/2003/05/071}{JHEP~0305,~071~(2003)}},
\texttt{\arxivref{hep-th/0304234}{hep-th/0304234}}.

\bibitem{Israel:2004vv}
D.~Israel, C.~Kounnas, D.~Orlando and P.~M.~Petropoulos,
\textit{``{Electric/magnetic deformations of S**3 and AdS(3), and geometric
  cosets}''},
\textsf{\doiref{10.1002/prop.200410190}{Fortsch.~Phys.~53,~73~(2005)}},
\texttt{\arxivref{hep-th/0405213}{hep-th/0405213}}.

\bibitem{Israel:2004cd}
D.~Israel, C.~Kounnas, D.~Orlando and P.~M.~Petropoulos,
\textit{``{Heterotic strings on homogeneous spaces}''},
\textsf{\doiref{10.1002/prop.200510250}{Fortsch.~Phys.~53,~1030~(2005)}},
\texttt{\arxivref{hep-th/0412220}{hep-th/0412220}}.

\bibitem{Orlando:2006cc}
D.~Orlando,
\textit{``{String Theory: Exact solutions, marginal deformations and hyperbolic
  spaces}''},
\textsf{\doiref{10.1002/prop.200610333}{Fortsch.~Phys.~55,~161~(2007)}},
\texttt{\arxivref{hep-th/0610284}{hep-th/0610284}}.

\bibitem{Detournay:2005fz}
S.~Detournay, D.~Orlando, P.~M.~Petropoulos and P.~Spindel,
\textit{``{Three-dimensional black holes from deformed anti-de Sitter}''},
\textsf{\doiref{10.1088/1126-6708/2005/07/072}{JHEP~0507,~072~(2005)}},
\texttt{\arxivref{hep-th/0504231}{hep-th/0504231}}.

\bibitem{Fredenhagen:2007rx}
S.~Fredenhagen, M.~R.~Gaberdiel and C.~A.~Keller,
\textit{``{Symmetries of perturbed conformal field theories}''},
\textsf{\doiref{10.1088/1751-8113/40/45/012}{J.~Phys.~A40,~13685~(2007)}},
\texttt{\arxivref{0707.2511}{arxiv:0707.2511}}.

\bibitem{Henningson:1992rn}
M.~Henningson and C.~R.~Nappi,
\textit{``{Duality, marginal perturbations and gauging}''},
\textsf{\doiref{10.1103/PhysRevD.48.861}{Phys.~Rev.~D48,~861~(1993)}},
\texttt{\arxivref{hep-th/9301005}{hep-th/9301005}}.

\bibitem{Lunin:2005jy}
O.~Lunin and J.~M.~Maldacena,
\textit{``{Deforming field theories with $U(1) \times U(1)$ global symmetry and
  their gravity duals}''},
\textsf{\doiref{10.1088/1126-6708/2005/05/033}{JHEP~0505,~033~(2005)}},
\texttt{\arxivref{hep-th/0502086}{hep-th/0502086}}.

\bibitem{Frolov:2005ty}
S.~A.~Frolov, R.~Roiban and A.~A.~Tseytlin,
\textit{``{Gauge-string duality for superconformal deformations of N=4 super
  Yang-Mills theory}''},
\textsf{\doiref{10.1088/1126-6708/2005/07/045}{JHEP~0507,~045~(2005)}},
\texttt{\arxivref{hep-th/0503192}{hep-th/0503192}}.

\bibitem{Frolov:2005dj}
S.~Frolov,
\textit{``{Lax pair for strings in Lunin-Maldacena background}''},
\textsf{\doiref{10.1088/1126-6708/2005/05/069}{JHEP~0505,~069~(2005)}},
\texttt{\arxivref{hep-th/0503201}{hep-th/0503201}}.

\bibitem{Alday:2005ww}
L.~F.~Alday, G.~Arutyunov and S.~Frolov,
\textit{``{Green-Schwarz strings in TsT-transformed backgrounds}''},
\textsf{\doiref{10.1088/1126-6708/2006/06/018}{JHEP~0606,~018~(2006)}},
\texttt{\arxivref{hep-th/0512253}{hep-th/0512253}}.

\bibitem{Klimcik:2008eq}
C.~Klimcik,
\textit{``{On integrability of the Yang-Baxter sigma-model}''},
\textsf{\doiref{10.1063/1.3116242}{J.Math.Phys.~50,~043508~(2009)}},
\texttt{\arxivref{0802.3518}{arxiv:0802.3518}}.

\bibitem{Delduc:2013qra}
F.~Delduc, M.~Magro and B.~Vicedo,
\textit{``{An integrable deformation of the AdS$_5 \times$S$^5$ superstring
  action}''},
\textsf{\doiref{10.1103/PhysRevLett.112.051601}{Phys.Rev.Lett.~112,~051601~(2014)}},
\texttt{\arxivref{1309.5850}{arxiv:1309.5850}}.

\bibitem{Kawaguchi:2014qwa}
I.~Kawaguchi, T.~Matsumoto and K.~Yoshida,
\textit{``{Jordanian deformations of the $AdS_5 x S^5$ superstring}''},
\textsf{\doiref{10.1007/JHEP04(2014)153}{JHEP~1404,~153~(2014)}},
\texttt{\arxivref{1401.4855}{arxiv:1401.4855}}.

\bibitem{Matsumoto:2014gwa}
T.~Matsumoto and K.~Yoshida,
\textit{``{Integrability of classical strings dual for noncommutative gauge
  theories}''},
\textsf{\doiref{10.1007/JHEP06(2014)163}{JHEP~1406,~163~(2014)}},
\texttt{\arxivref{1404.3657}{arxiv:1404.3657}}.

\bibitem{vanTongeren:2015uha}
S.~J.~van~Tongeren,
\textit{``{Yang–Baxter deformations, AdS/CFT, and twist-noncommutative gauge
  theory}''},
\textsf{\doiref{10.1016/j.nuclphysb.2016.01.012}{Nucl.~Phys.~B904,~148~(2016)}},
\texttt{\arxivref{1506.01023}{arxiv:1506.01023}}.

\bibitem{vanTongeren:2016eeb}
S.~J.~van~Tongeren,
\textit{``{Almost abelian twists and AdS/CFT}''},
\textsf{\doiref{10.1016/j.physletb.2016.12.002}{Phys.~Lett.~B765,~344~(2017)}},
\texttt{\arxivref{1610.05677}{arxiv:1610.05677}}.

\bibitem{Araujo:2017jkb}
T.~Araujo, I.~Bakhmatov, E.~O.~Colg\'{a}in, J.~Sakamoto, M.~M.~Sheikh-Jabbari
  and K.~Yoshida,
\textit{``{Yang-Baxter $\sigma$-models, conformal twists, and noncommutative
  Yang-Mills theory}''},
\textsf{\doiref{10.1103/PhysRevD.95.105006}{Phys.~Rev.~D95,~105006~(2017)}},
\texttt{\arxivref{1702.02861}{arxiv:1702.02861}}.

\bibitem{Osten:2016dvf}
D.~Osten and S.~J.~van~Tongeren,
\textit{``{Abelian Yang–Baxter deformations and TsT transformations}''},
\textsf{\doiref{10.1016/j.nuclphysb.2016.12.007}{Nucl.~Phys.~B915,~184~(2017)}},
\texttt{\arxivref{1608.08504}{arxiv:1608.08504}}.

\bibitem{Hoare:2016wsk}
B.~Hoare and A.~A.~Tseytlin,
\textit{``{Homogeneous Yang-Baxter deformations as non-abelian duals of the
  AdS$_5$ sigma-model}''},
\textsf{\doiref{10.1088/1751-8113/49/49/494001}{J.~Phys.~A49,~494001~(2016)}},
\texttt{\arxivref{1609.02550}{arxiv:1609.02550}}.

\bibitem{Borsato:2016pas}
R.~Borsato and L.~Wulff,
\textit{``{Integrable Deformations of $T$-Dual $\sigma$ Models}''},
\textsf{\doiref{10.1103/PhysRevLett.117.251602}{Phys.~Rev.~Lett.~117,~251602~(2016)}},
\texttt{\arxivref{1609.09834}{arxiv:1609.09834}}.

\bibitem{Lust:2018jsx}
D.~L{\"u}st and D.~Osten,
\textit{``{Generalised fluxes, Yang-Baxter deformations and the $O(d,d)$
  structure of non-abelian T-duality}''},
\textsf{\doiref{10.1007/JHEP05(2018)165}{JHEP~1805,~165~(2018)}},
\texttt{\arxivref{1803.03971}{arxiv:1803.03971}}.

\bibitem{Sakamoto:2018krs}
J.-i.~Sakamoto and Y.~Sakatani,
\textit{``{Local $\beta$-deformations and Yang-Baxter sigma model}''},
\texttt{\arxivref{1803.05903}{arxiv:1803.05903}}.

\bibitem{Borsato:2018idb}
R.~Borsato and L.~Wulff,
\textit{``{Non-abelian T-duality and Yang-Baxter deformations of Green-Schwarz
  strings}''},
\textsf{\doiref{10.1007/JHEP08(2018)027}{JHEP~1808,~027~(2018)}},
\texttt{\arxivref{1806.04083}{arxiv:1806.04083}}.

\bibitem{Bakhmatov:2017joy}
I.~Bakhmatov, {\"O}.~Kelekci, E.~O.~Colg\'{a}in and M.~M.~Sheikh-Jabbari,
\textit{``{Classical Yang-Baxter Equation from Supergravity}''},
\texttt{\arxivref{1710.06784}{arxiv:1710.06784}}.

\bibitem{Alvarez:1994np}
E.~Alvarez, L.~Alvarez-Gaume and Y.~Lozano,
\textit{``{On nonAbelian duality}''},
\textsf{\doiref{10.1016/0550-3213(94)90093-0}{Nucl.~Phys.~B424,~155~(1994)}},
\texttt{\arxivref{hep-th/9403155}{hep-th/9403155}}.

\bibitem{Elitzur:1994ri}
S.~Elitzur, A.~Giveon, E.~Rabinovici, A.~Schwimmer and G.~Veneziano,
\textit{``{Remarks on nonAbelian duality}''},
\textsf{\doiref{10.1016/0550-3213(94)00426-F}{Nucl.~Phys.~B435,~147~(1995)}},
\texttt{\arxivref{hep-th/9409011}{hep-th/9409011}}.

\bibitem{Arutyunov:2015mqj}
G.~Arutyunov, S.~Frolov, B.~Hoare, R.~Roiban and A.~A.~Tseytlin,
\textit{``{Scale invariance of the $\eta$-deformed $AdS_5\times S^5$
  superstring, T-duality and modified type II equations}''},
\textsf{\doiref{10.1016/j.nuclphysb.2015.12.012}{Nucl.~Phys.~B903,~262~(2016)}},
\texttt{\arxivref{1511.05795}{arxiv:1511.05795}}.

\bibitem{Wulff:2016tju}
L.~Wulff and A.~A.~Tseytlin,
\textit{``{Kappa-symmetry of superstring sigma model and generalized 10d
  supergravity equations}''},
\textsf{\doiref{10.1007/JHEP06(2016)174}{JHEP~1606,~174~(2016)}},
\texttt{\arxivref{1605.04884}{arxiv:1605.04884}}.

\bibitem{Wulff:2018aku}
L.~Wulff,
\textit{``{Trivial solutions of generalized supergravity vs non-abelian
  T-duality anomaly}''},
\textsf{\doiref{10.1016/j.physletb.2018.04.025}{Phys.~Lett.~B781,~417~(2018)}},
\texttt{\arxivref{1803.07391}{arxiv:1803.07391}}.

\bibitem{Borsato:2016ose}
R.~Borsato and L.~Wulff,
\textit{``{Target space supergeometry of $\eta$ and $\lambda$-deformed
  strings}''},
\textsf{\doiref{10.1007/JHEP10(2016)045}{JHEP~1610,~045~(2016)}},
\texttt{\arxivref{1608.03570}{arxiv:1608.03570}}.

\bibitem{Araujo:2018rho}
T.~Araujo, E.~Ã.~Colgáin, Y.~Sakatani, M.~M.~Sheikh-Jabbari and H.~Yavartanoo,
\textit{``{Holographic integration of $T \bar{T}$ \& $J \bar{T}$ via
  $O(d,d)$}''},
\texttt{\arxivref{1811.03050}{arxiv:1811.03050}}.

\bibitem{Giveon:2017nie}
A.~Giveon, N.~Itzhaki and D.~Kutasov,
\textit{``{$ \mathrm{T}\overline{\mathrm{T}} $ and LST}''},
\textsf{\doiref{10.1007/JHEP07(2017)122}{JHEP~1707,~122~(2017)}},
\texttt{\arxivref{1701.05576}{arxiv:1701.05576}}.

\bibitem{Forste:1994wp}
S.~Forste,
\textit{``{A Truly marginal deformation of SL(2, R) in a null direction}''},
\textsf{\doiref{10.1016/0370-2693(94)91340-4}{Phys.~Lett.~B338,~36~(1994)}},
\texttt{\arxivref{hep-th/9407198}{hep-th/9407198}}.

\bibitem{Israel:2003ry}
D.~Israel, C.~Kounnas and M.~P.~Petropoulos,
\textit{``{Superstrings on NS5 backgrounds, deformed AdS(3) and holography}''},
\textsf{\doiref{10.1088/1126-6708/2003/10/028}{JHEP~0310,~028~(2003)}},
\texttt{\arxivref{hep-th/0306053}{hep-th/0306053}}.

\bibitem{Chakraborty:2018vja}
S.~Chakraborty, A.~Giveon and D.~Kutasov,
\textit{``{$ J\overline{T} $ deformed CFT$_{2}$ and string theory}''},
\textsf{\doiref{10.1007/JHEP10(2018)057}{JHEP~1810,~057~(2018)}},
\texttt{\arxivref{1806.09667}{arxiv:1806.09667}}.

\bibitem{Apolo:2018qpq}
L.~Apolo and W.~Song,
\textit{``{Strings on warped AdS$_{3}$ via $ \mathrm{T}\bar{\mathrm{J}} $
  deformations}''},
\textsf{\doiref{10.1007/JHEP10(2018)165}{JHEP~1810,~165~(2018)}},
\texttt{\arxivref{1806.10127}{arxiv:1806.10127}}.

\bibitem{Cagnazzo:2012se}
A.~Cagnazzo and K.~Zarembo,
\textit{``{B-field in AdS$_3$/CFT$_2$ Correspondence and Integrability}''},
\textsf{\doiref{10.1007/JHEP11(2012)133,
  10.1007/JHEP04(2013)003}{JHEP~1211,~133~(2012)}},
\texttt{\arxivref{1209.4049}{arxiv:1209.4049}}.

\bibitem{Sundin:2012gc}
P.~Sundin and L.~Wulff,
\textit{``{Classical integrability and quantum aspects of the AdS(3) x S(3) x
  S(3) x S(1) superstring}''},
\textsf{\doiref{10.1007/JHEP10(2012)109}{JHEP~1210,~109~(2012)}},
\texttt{\arxivref{1207.5531}{arxiv:1207.5531}}.

\bibitem{Azeyanagi:2012zd}
T.~Azeyanagi, D.~M.~Hofman, W.~Song and A.~Strominger,
\textit{``{The Spectrum of Strings on Warped $AdS_3 x S^3$}''},
\textsf{\doiref{10.1007/JHEP04(2013)078}{JHEP~1304,~078~(2013)}},
\texttt{\arxivref{1207.5050}{arxiv:1207.5050}}.

\bibitem{Delduc:2014uaa}
F.~Delduc, M.~Magro and B.~Vicedo,
\textit{``{Integrable double deformation of the principal chiral model}''},
\textsf{\doiref{10.1016/j.nuclphysb.2014.12.018}{Nucl.~Phys.~B891,~312~(2015)}},
\texttt{\arxivref{1410.8066}{arxiv:1410.8066}}.

\bibitem{Delduc:2017fib}
F.~Delduc, B.~Hoare, T.~Kameyama and M.~Magro,
\textit{``{Combining the bi-Yang-Baxter deformation, the Wess-Zumino term and
  TsT transformations in one integrable $\sigma$-model}''},
\textsf{\doiref{10.1007/JHEP10(2017)212}{JHEP~1710,~212~(2017)}},
\texttt{\arxivref{1707.08371}{arxiv:1707.08371}}.

\bibitem{Delduc:2018xug}
F.~Delduc, B.~Hoare, T.~Kameyama, S.~Lacroix and M.~Magro,
\textit{``{Three-parameter integrable deformation of $\mathbb{Z}_4$ permutation
  supercosets}''},
\texttt{\arxivref{1811.00453}{arxiv:1811.00453}}.

\bibitem{Klimcik:1995dy}
C.~Klimcik and P.~Severa,
\textit{``{Poisson-Lie T duality and loop groups of Drinfeld doubles}''},
\textsf{\doiref{10.1016/0370-2693(96)00025-1}{Phys.Lett.~B372,~65~(1996)}},
\texttt{\arxivref{hep-th/9512040}{hep-th/9512040}}.

\bibitem{Stern:1998my}
A.~Stern,
\textit{``{Hamiltonian approach to Poisson Lie T - duality}''},
\textsf{\doiref{10.1016/S0370-2693(99)00111-2}{Phys.~Lett.~B450,~141~(1999)}},
\texttt{\arxivref{hep-th/9811256}{hep-th/9811256}}.

\bibitem{Klimcik:2015gba}
C.~Klimcik,
\textit{``{$\eta$ and $\lambda$ deformations as ${\cal E}$-models}''},
\textsf{\doiref{10.1016/j.nuclphysb.2015.09.011}{Nucl.~Phys.~B900,~259~(2015)}},
\texttt{\arxivref{1508.05832}{arxiv:1508.05832}}.

\bibitem{Demulder:2018lmj}
S.~Demulder, F.~Hassler and D.~C.~Thompson,
\textit{``{Doubled aspects of generalised dualities and integrable
  deformations}''},
\texttt{\arxivref{1810.11446}{arxiv:1810.11446}}.

\bibitem{Smirnov:2016lqw}
F.~A.~Smirnov and A.~B.~Zamolodchikov,
\textit{``{On space of integrable quantum field theories}''},
\textsf{\doiref{10.1016/j.nuclphysb.2016.12.014}{Nucl.~Phys.~B915,~363~(2017)}},
\texttt{\arxivref{1608.05499}{arxiv:1608.05499}}.

\bibitem{Cavaglia:2016oda}
A.~Cavagli\`{a}, S.~Negro, I.~M.~Sz\'ecs\'enyi and R.~Tateo,
\textit{``{$T \bar{T}$-deformed 2D Quantum Field Theories}''},
\textsf{\doiref{10.1007/JHEP10(2016)112}{JHEP~1610,~112~(2016)}},
\texttt{\arxivref{1608.05534}{arxiv:1608.05534}}.

\bibitem{Guica:2017lia}
M.~Guica,
\textit{``{An integrable Lorentz-breaking deformation of two-dimensional
  CFTs}''},
\textsf{\doiref{10.21468/SciPostPhys.5.5.048}{SciPost~Phys.~5,~048~(2018)}},
\texttt{\arxivref{1710.08415}{arxiv:1710.08415}}.

\bibitem{STOLIN1999285}
A.~Stolin,
\textit{``Rational solutions of the classical Yang-Baxter equation and quasi
  Frobenius Lie algebras''},
\textsf{\doiref{http://dx.doi.org/10.1016/S0022-4049(97)00217-X}{Journal~of~Pure~and~Applied~Algebra~137,~285
  ~(1999)}},
\href{http://www.sciencedirect.com/science/article/pii/S002240499700217X}{\texttt{http://www.sciencedirect.com/science/article/pii/S002240499700217X}}.

\bibitem{zbMATH03411826}
B.-Y.~{Chu},
\textit{``Symplectic homogeneous spaces''},
\textsf{Trans.~Am.~Math.~Soc.~197,~145~(1974)}.

\bibitem{Ovando2006}
G.~Ovando,
\textit{``Four dimensional symplectic Lie algebras.''},
\textsf{Beiträge~zur~Algebra~und~Geometrie~47,~419~(2006)},
\href{http://eudml.org/doc/226718}{\texttt{http://eudml.org/doc/226718}}.

\bibitem{Stolin:1991}
A.~Stolin,
\textit{``Constant solutions of Yang-Baxter equation for $\mathfrak{sl}(2)$ and
  $\mathfrak{sl}(3)$''},
\textsf{Mathematica~Scandinavica~69,~81~(1991)},
\href{http://www.jstor.org/stable/24492601}{\texttt{http://www.jstor.org/stable/24492601}}.

\bibitem{Burde:2016}
D.~Burde and W.~A.~Moens,
\textit{``Commutative post-Lie algebra structures on Lie algebras''},
\textsf{\doiref{https://doi.org/10.1016/j.jalgebra.2016.07.030}{Journal~of~Algebra~467,~183
  ~(2016)}},
\href{http://www.sciencedirect.com/science/article/pii/S002186931630240X}{\texttt{http://www.sciencedirect.com/science/article/pii/S002186931630240X}}.

\end{thebibliography}

\end{document}